\documentclass[usenatbib,a4paper]{mn2e}
\usepackage{graphicx,amsmath}
\usepackage{url}
\usepackage[dvipsnames,hyperref]{xcolor}
\usepackage[colorlinks=true,linkcolor=Black,citecolor=ForestGreen,urlcolor=Blue]{hyperref}
\usepackage{txfonts}
\bibpunct{(}{)}{;}{a}{}{,}

\makeatletter
\newcommand{\boxedr}[1]{\textcolor{red}{%
  \fbox{\normalcolor\m@th$\displaystyle#1$}}}
\newcommand{\boxedg}[1]{\textcolor{green}{%
  \fbox{\normalcolor\m@th$\displaystyle#1$}}}
\makeatother

\newcommand\oh{\frac{1}{2}}
\newcommand\os{\frac{1}{6}}
\newcommand\ov{\frac{1}{48}}
\newcommand\ow{\frac{2}{48}}
\newcommand\oz{\frac{4}{48}}
\newcommand\op{\frac{1}{3}e^{i2\pi/3}}
\newcommand\om{\frac{1}{3}e^{-i2\pi/3}}

\newcommand\oa{\frac{1}{24}}
\newcommand\ob{\frac{1}{24}e^{i\pi}}
\newcommand\oc{\frac{1}{24}e^{i\pi/2}}
\newcommand\od{\frac{1}{24}e^{i3\pi/2}}

\newcommand\gop{\frac{1}{12}}
\newcommand\ggoa{\frac{1}{6}}
\newcommand\gob{\frac{1}{6}e^{i\pi}}
\newcommand\goc{\frac{1}{6}e^{i\pi/2}}
\newcommand\god{\frac{1}{6}e^{i3\pi/2}}
   
\title{A new interferometer architecture combining nulling with phase closure measurements}
\author[S. Lacour et al.]{S. Lacour,$^1$ P. Tuthill,$^2$ J. D. Monnier,$^3$ T. Kotani,$^4$ L. Gauchet$^1$ and P.
Labeye$^5$\\
$^{1}$  LESIA/Observatoire de Paris, CNRS, UPMC, Universit\'e Paris Diderot, 5 place Jules Janssen, 92195 Meudon,
France \\
$^{2}$ Sydney Institute for Astronomy, School of Physics, The Univeristy of Sydney, N.S.W. 2006, Australia \\
$^{3}$ Department of Astronomy, University of Michigan, 941 Dennison Building, Ann Arbor, MI 48109, USA  \\
$^{4}$  National Astronomical Observatory of Japan, 2-21-1 Osawa, Mitaka, Tokyo 181-8588, Japan  \\
$^{5}$  CEA-LETI, MINATEC Campus, 17 rue des Martyrs, 38054 Grenoble Cedex 9, France  \\}

\begin{document}

\label{firstpage}

\maketitle

\begin{abstract}

Imaging the direct light signal from a faint exoplanet against the overwhelming glare of its host
star presents one of the fundamental challenges to modern astronomical instrumentation.
Achieving sufficient signal-to-noise for detection by direct imaging is limited by three basic
physical processes: aberration of the wavefronts (both instrumental and atmospheric), photon noise,
and detector noise. 
In this paper, we advance a novel optical setup which synthesizes the advantages of two different 
techniques: nulling interferometry to mitigate photon noise, and closure phase to combat optical aberrations.
Our design, which employs technology from integrated optics and photonics, is intended to combine the 
advantageous aspects of both a coronagraph and a non-redundant interferometer inside a single optical device. 
We show that such an instrument would have a dynamic range limited either by i) the readout noise (if perfect co-phasing), or
 ii) the photon noise due to stellar flux leakage (in the case of imperfect nulling). 
This concept is optimal  when the readout noise is not the main limitation, ie, for space interferometry or for ground based observations of 
bright stellar hosts (apparent magnitude brighter than 10).
\end{abstract}

\begin{keywords}
 Exoplanets detection -- nulling -- interferometry
\end{keywords}

%
%________________________________________________________________

\section{Introduction}

The problem of obtaining high dynamic range imaging is encountered throughout astronomy, as often critical but 
faint physical processes are played out on a stage so remote so as to be buried in the glare of a bright stellar
core. 
The advent of exoplanetery discovery and characterization has added dramatic impetus to the solution of this
challenge, which is fundamentally limited by three basic noise processes with different statistical properties: 
\begin{itemize}
 \item {\it Phase Noise}, also called speckle noise, arises from imperfect transfer of the wavefronts through the 
atmosphere and the instrument. 
This encompasses any optical perturbation that degrades the optical transfer function from ideal (e.g. the Airy pattern 
for a full circular aperture). 
Because phase noise redistributes the flux of the astronomical object in the image plane, it limits the dynamic range 
to a value which is independent of the brightness of the source.
\item {\it Photon Noise} is proportional to the variance the number of photons. 
By diverting the flux from a bright on-axis stellar host away from the science camera, coronagraph designs are 
effective also in suppressing the attendant photon noise from the host star, and thereby increasing the signal-to-noise
for detection of an off-axis exoplanet.
In the photon-noise limited regime, the dynamic range increases with the square root of the brightness. 
\item {\it Readout Noise}, including dark current, has a fixed value which does not depend on the brightness of the source. 
In the readout-noise limited regime, the dynamic range obtained increases linearly as a function of the 
brightness of the target.
\end{itemize}
Of course, by considering only these three noises, we considerably simplify the problem of high dynamic range imaging.
There are many sources of error and mis-calibrations such as blurring as a result of long integrations (compared to the 
coherence time), the effects of detector non-linearity, chromatic dispersion, bad pixels, field distortions, imperfect 
treatment of polarization and many more which must be dealt with or accounted for separately.

But the best we can hope for is that the ultimate dynamic range of an imaging observation is limited by whichever one of 
the three {\it fundamental} noises predominates. 
Ideally, we want our experiment to operate in the detector noise limited regime (in which phase and photon noise are
so well controlled as to be minor contributors) for which the dynamic range obtained is fixed
for any given integration time. 
There are numerous approaches to mitigating photon and phase noise.
For the case of photon noise from the bright host star masking the exoplanetary signal, the most direct approach is to 
spatially separate star and planetary light, for example by increasing the angular resolution of the telescope. 
Specific instrumental setups which combat photon noise are coronagraphs and nulling interferometers. 
In fact these latter two devices are intimately related, and the operating principle of the 4-quadrant phase mask
coronagraph \citep{2000PASP..112.1479R} is highly analogous to that of a two-telescope interferometer nuller 
\citep{1978Natur.274..780B}: both techniques use a  $\pi$ delay on half of the stellar light to extinguish the
source on-axis. To address the phase noise, the most direct approach is real time correction of phase aberrations by
means of an adaptive optics system. However, post-processing techniques have also proven valuable. 
For example, Speckle interferometry \citep{1974ApJ...194L.147L} has been shown to be highly successful, as well as new 
methods like Kernel phase \citep{2010ApJ...724..464M}, which is an extension of the well-established closure phases.

However it turns out that post-processing techniques such as Kernel phase are difficult to adapt to coronagraphic datasets \citep{2007lyot.confE..40L}. The assumptions underlying Kernel phase are robust against small phase errors introduced by an instrument, but
not against the major apodization and wavefront amplitude changes which occur in a coronagraph.
As a result, the ``Kernel space'' of the phases no longer exists, and the theory has to be recast. The philosophy of this paper is to propose a single optical 
system to optimize detection of high contrast companions in the presence of both photon and wavefront phase noise. 
Our proposition is to optically combine, for the first time, a nulling interferometer (to combat photon noise) together
with a clasical 3 telescopes beam combiner (to make use of the self-calibrating closure phase observable -- resilient against phase noise -- commonly employed in 
optical interferometry). \footnote{Note that this optical system must not be mistaken with the concept of {\it phase closure nulling} \citep{2010A&A...509A..66D} or {\it closure-phase nulling Interferometer} \citep{2006ApJ...645.1554D}. These two papers propose reduction techniques using the closure phase observable on traditional optical setups (a three telescope interferometer on one hand, and a three telescope nuller on the other hand). On the contrary here, we are proposing an optical architecture to enhance the signal of an extrasollar planet on the closure phase observable.}

The mathematical foundation underlying the concept is presented in Section~\ref{sc:concept}. 
Next, we sketch a realized instrument using this principle based on single-mode filtering and integrated optics. 
Our analog of a coronagraphic phase mask is performed by extinguishing the light in tri-couplers that we call nullers. 
These are presented Section~\ref{sc:nul}. 
The second stage of the detection process is achieved by combining the light from each of the nulled outputs to produce 
fringes and to allow measurement of a closure-phase type observable. 
Such a combiner is presented Section~\ref{sc:bc} and simulated in Section~\ref{sc:res}. 
Finally, section~\ref{sc:con} discusses possible applications and extensions of this concept.

\section{Mathematical Basis}
\label{sc:concept}

\subsection{Coherence and nulling}
\label{sc:basic}

The coherence of two light beams is defined as:
\begin{equation}
C_{12}=<E_1E_2^*>
\end{equation}
where $E_1$ and $E_2$ are complex phasor representations of light beams such that the instantaneous electric field at some location labeled "1" at time $t$ is given by:
\begin{equation}
\psi_1=\Re[E_1\exp(2 \pi i \nu t)]
\end{equation}
and $<>$ denotes averaging over the detector integration time. For a quasi-monochromatic beam ($\nu$ is the wavenumber), it is important to note that $E_1$ and $E_2$ are randomly-varying quantities in both amplitude and phase, and that these random variations take place on timescales comparable to the coherence time of the radiation, which is assumed to be very short compared to the detector integration time but long compared to 1/$\nu$.

To take the simples example of spatial coherence, we can consider two point collectors of light at locations 1 and 2 which transmit light via monomode optical fibers to the point at which point the waves are interfered. The fibres will both delay the beams and lose some light, and the effect of
the fibres can be represented by a complex “antenna gain” $G$ for each fibre.
The gain is stable over periods long compared with the detector integration
time, meaning that for short integration time, $G$ can also encompass the atmospheric amplitude and phase errors. 
If the phasors at the entrance of the fibres are $E_1$ and $E_2$ then at the
exits of the fibres they are $G_1 E_1$ and $G_2 E_2$ . The interference pattern at the
exit of the fibres will therefore have terms which depend on $G_1 G_2^*
<E_1 E_2^*>$. Nulling implies control of $G_1$ and $G_2$ so as to ensure destructive interference
on a point source.

To take the simplest example of an interesting source, we can consider a binary 
system where the primary is located directly over the pair of collectors,
such that the illumination of the primary (A) arrives simultaneously at both
collectors, but the secondary (B) arrives slightly out of phase between the
apertures due to the angular offset and the baseline. We can write this
situation as
\begin{equation}
E_1 = E_A + E_B
\end{equation}
and
\begin{equation}
E_2 = E_A + \exp(i\theta)E_B
\end{equation}
where $E_A$ and $E_B$ are phasors representing mutually uncorrelated fields ar-
riving from the primary and secondary respectively. In this case we have
\begin{equation}
E_1 E_2^*
= |E_A |^2 + |E_B |^2 \exp(-i\theta) + E_A E_B^*
\exp(-i\theta) + E_A^*
E_B
\end{equation}
All the terms in this expansion of $E_1 E_2^*$ are randomly-varying on timescales
of the coherence time. In a simplified world where only the phases of $E_A$ and
$E_B$ vary, then only the last two terms are random, but nevertheless $E_1E_2^*$
describes a random quantity, and cannot be simply related to $<E_1E_2^*>$, 
which is a constant for a given object and baseline vector.
The first two terms above have non-zero means when averaged over the
response time of the detector while the latter two are zero-mean ($<E_A E_B^*>=0$) . Thus the
coherence detected at the output of the fibres is of the form
\begin{eqnarray}
C_{12}&=&G_1 G^*_2 <E_1 E_2^*> \label{eq:C12a} \\
&=& G_1 G_2^* \left(<|E_A |^2> + <|E_B |^2 >\exp(-i\theta)\right)
\end{eqnarray}
which can be related to the complex visibility of the binary star $V_{12}$ as
\begin{equation}
C_{12}=G_1 G^*_2  V_{12} \left(<|E_A |^2> + <|E_B |^2>\right)
\label{eq:C12bi}
\end{equation}
with
 \begin{equation}
V_{12} =\cfrac{<|E_A |^2> + <|E_B |^2 >\exp(-i\theta) }{ <|E_A |^2> + <|E_B |^2 >}\,.
\end{equation}
This complex visibility value is the normalized Fourier transform of the binary object. More importantly, the Van Cittert-Zernike theorem generalize that demonstration to any type of object. Hence,  Eq.~(\ref{eq:C12bi}) can be generalized to:
\begin{equation}
C_{12}=G_1 G^*_2  V_{12}
\label{eq:C12b}
\end{equation}
where $V_{12}$ is the complex Fourier transform of the brightness distribution, and $C_{12}$ the coherence value after normalization by the total flux of the astronomical object.

The closure phase is then often calculated directly from the argument of the bispectrum:
\begin{equation}
 \Phi_{\rm CP} = \arg{(B_{123} )}
\end{equation}
where $B_{123}$ is a multiplication of the coherence terms forming a triangle:
\begin{equation}
B_{123} =C_{12} \cdot C_{23} \cdot C_{13}^*\,.\label{eq:CPorig} 
\end{equation}

\subsection{Closure phase on the combined nulled output}

   \begin{figure}
   \centering
   \includegraphics[scale=.7]{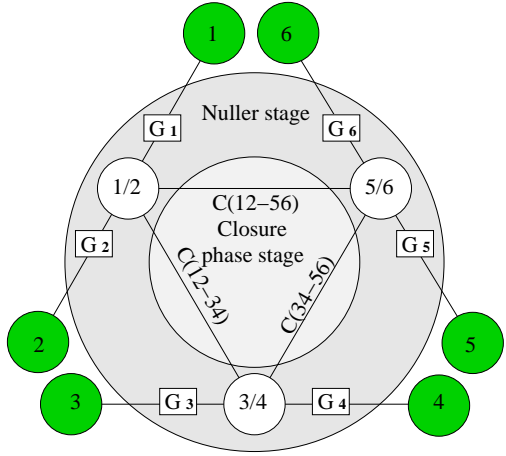} \vspace{0.1cm} 
      \caption{Schematic representation of a 6-beam closure phase nuller. 
The input beams are represented here by the dark green circles. 
In the first (nulling) stage of the instrument, three nulled outputs are obtained: 
Beam 1 is delayed by $\pi$ and is combined with beam 2 to give the null (in the circle 1/2). 
Similarly beams 3 and 4, and beams 5 and 6 complete the triangle at the first nulling stage. 
In the second stage, interferometric recombination of these three fields yield coherence values 
which can be used to constraint the visibilities.}
         \label{fig:principle}
   \end{figure}

To combine the functionality of nulling with closure phase, the interferometer must operates on the following two basic principles.
First, several beams must be combined pair-wise, while phase tracking is employed to produce an
output for each which lies at an interferometric null. 
Second, the outputs of these beams must be combined with each other to retrieve the complex coherence of the light. 
Complex visibilities are extracted, and closure phase can be calculated.
The closure phases should be equal to zero if the source is spatially coherent 
(such as a simple un-resolved point source). 
On the contrary, if the source is not fully coherent (for example, due to the presence of a planet), 
the closure phase will deviate from zero. The main idea is that the null filters the majority of the stellar flux, 
enhancing the flux ratio between star to planet as well as the increasing the value of the closure phases.

Such an instrument is presented in a schematic view in Fig.~\ref{fig:principle} where we sketch a setup to
combine 6 beams (represented as the green circles). Beam 1 is delayed and combined with beam 2 to give a
null output in the circle (1/2). 
The same is done with beam 3 and beam 4, and with beams 5 and 6. 
These 3 nulled outputs are then combined pair-wise to extract three coherency values: 
$C_{12-34}$,  $C_{34-56}$ and $C_{12-56}$. 
The coherency values are now different from the one detailed in the previous section. They are obtained from a combination of four electric fields:
\begin{equation}
C_{12-34}=<(G_1 E_1+G_2 E_2)\cdot(G_3 E_3+G_4 E_4)^*>
\end{equation}
where $G_1$ and $G_2$ are the antenna gains to ensure destructive interference between $E_1$ and $E_2$, as are $G_3$ and $G_4$ to ensure destruction of $E_3$ and $E_4$.

Following the same approach as the one performed in section~\ref{sc:basic}, we can link these coherency values to the theoretical coherence of the light on the entrance beams:
\begin{eqnarray}
C_{12-34}&=&G_1 G^*_3 <E_1 E_3^*> +G_1 G^*_4 <E_1 E_4^*> \nonumber \\
&& +G_2 G^*_3 <E_2 E_3^*> +G_2 G^*_4 <E_2 E_4^*>   \\
&=& C_{13} + C_{14} + C_{23} + C_{24} 
\label{eq:C1234decomp}
\end{eqnarray}
and to the spatial coherency values of the astronomical object:
\begin{equation}
C_{12-34} = G_1 G_3^* V_{13}+G_1 G_4^* V_{14}+G_2 G_3^* V_{23}+G_2 G_4^* V_{24}
\end{equation}
according to Eq.~(\ref{eq:C12b}).
 Interestingly enough, the spatial coherency of the baselines 12 and 34 does not appear in that relation.

Similarly to what is done in Section~\ref{sc:basic}, a quantity equivalent to the closure phase can be calculated:
\begin{equation}
 \Phi_{\rm CP null} = \arg{(B_{12-34-56} )}\label{eq:CP} \,.
\end{equation}
where $B_{12-34-56}$ is a multiplication of the coherence terms between three nuls:
\begin{equation}
B_{12-34-56} = \left(C_{12-34} \cdot  C_{34-56}  \cdot C_{12-56} ^*\right)  \label{eq:tri}
\end{equation}
Hence
 \small
\begin{eqnarray}
B_{12-34-56} &=&\left(   G_1G_3^*V_{13} + G_2G_4^*V_{24}+G_1G_4^*V_{14} + G_2G_3^*V_{23}\right) \nonumber \\ \nonumber
             &&\cdot \left(   G_3G_5^*V_{35} + G_4G_6^*V_{46} + G_3G_6^*V_{36} + G_4G_5^*V_{45}\right) \\
             &&\cdot \left(   G_1G_5^*V_{15} + G_2G_4^*V_{26} + G_1G_6^*V_{16} + G_2G_5^*V_{25}\right)^* 
\label{eq:triple}
\end{eqnarray}
\normalsize

 To appreciate the closure phase's value $\Phi_{\rm CP null}$, it is essential to realize that this particular
closure phase does not imply the closure of a triangle between three telescopes, but between three baselines: 1-2, 3-4 and 5-6.
On a traditional version of the bispectrum, an important property of the triple product is
that its values are equal to: 
\begin{eqnarray}
B_{123}&=&C_{12} \cdot C_{23} \cdot C_{13}^*\\
&=& G_{1}G_{2}^* V_{12} \cdot G_{2}G_{3}^* V_{23} \cdot G_{3}G_{1}^* V_{13}^* \\
&=& |G_{1}|^2|G_{2}|^2|G_{3}|^2 V_{12}V_{23}V_{13}^* \,.
\end{eqnarray}
As a consequence, for a point source ($V_{12}=V_{23}=V_{14}=1$), 
the bispectrum is real: the closure phase is nul. The
same property applies with the triple product calculated over a triangle of baselines if $V_{12}=V_{13}=V_{14}=V_{23}=V_{24}=V_{34}=1$ :
\begin{eqnarray}
B_{12-34-56}&=&\left(   G_1G_3^*   + G_2G_4^* + G_1G_4^* + G_2G_3^*\right) \nonumber \\ \nonumber
             &&\cdot \left(   G_3G_5^* + G_4G_6^* + G_3G_6^* + G_4G_5^*\right) \\
             &&\cdot \left(   G_1G_5^* + G_2G_4^* + G_1G_6^* + G_2G_5^*\right)^* \\
             &=&  (G_1+G_2) (G_3 + G_4)^* \nonumber \\ \nonumber
             &&\cdot  (G_3+G_4) (G_5 + G_6)^* \\
             &&\cdot \left(   (G_1+G_2) (G_5 + G_6)^*   \right)^* \\
&=&|G_{1}+G_{2}|^2|G_{3}+G_4|^2|G_5+G_6|^2 \,.
\end{eqnarray}

An important result of this demonstration is that if the beam is fully coherent, the closure phase obtained on three combined baselines is equally egal to zero: $\arg{(B_{12-34-56})}=0$. On the other hand, if one of the $V_{ij}$ is not equal to 1, then $\arg{(B_{12-34-56})}$ can be different from zero. The remaining issue is that, although $\Phi_{\rm CP null} \neq 0$ reveals that there
must be {\em some} departure from a simple unresolved source, it is not trivial to link it to a specific resolved structure. In other words, the $\Phi_{\rm CP null}$ are not straightforwardly related to the visibilities, as were the classical $\Phi_{\rm CP}$ . In the following section of the paper, we
demonstrate for a defined instrumental configuration that it is however possible to recover direct
information about the $V_{ij}$ from the coherency values $C_{ij-kl}$. 

\subsection{Relationship between spatial coherency and triple product}
\label{sc:triple}

   \begin{figure}
   \centering
   \includegraphics[scale=.7]{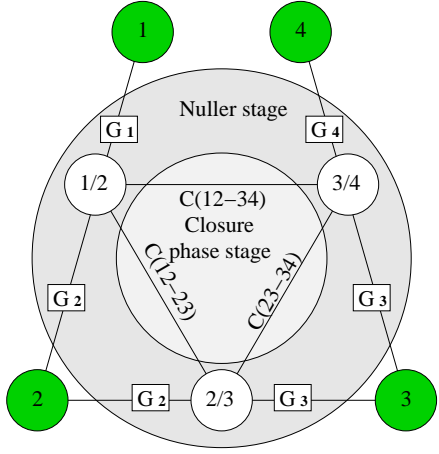} 
      \caption{Schematic representation of a 4 telescope closure phase nuller. The difference with respect to Fig.~\ref{fig:principle} is that the inputs 2 and 3, as well as 4 and 5, are merged into one. The goal is to cophase the whole array to produce the meaningful coherency values: 
$C_{12-23}$,  $C_{23-34}$ and $C_{12-34}$. These values can be used to constrain the complex visibilities $V_{1223}$,  $V_{2334}$ and $V_{1234}$. More specifically, the relation between the coherency values and the spatial coherence of the astronomical object is stated in Eq.~(\ref{eq:ImRel}).
}
         \label{fig:Nul}
   \end{figure}

The triple product combination is obtained in Eq.~\ref{eq:triple} thanks to
the derivation of the three coherence measurements $C_{12-34}$,  $C_{34-56}$,  $C_{56-12}$. Simplification of this equation is hard without any approximation. Fortunatly, we can assume that {\it some} nulling is achieved. The nulls are maintained by controlling the electric field of each beam prior to the coupler. To that end, the antenna gains must be pairwise of equal amplitude, but in phase opposition:
\begin{eqnarray}
G_2&=&G_1 (\epsilon_{2} -1) \\
G_4&=&G_3(\epsilon_{4}-1) \\
G_6&=&G_5(\epsilon_{6}-1) 
\end{eqnarray}
Of course, the better the phase tracking, the smaller the $\epsilon_{i}$ and the better the nulling. Concerning the spatial coherency values, we will similarly assume that the object is close to having a point-like type of structure: 
\begin{eqnarray}
V_{ij}&=&1+v_{ij} 
\end{eqnarray}
with $v_{ij}$ small. Hence, we can decompose the coherence vectors into three quantities of decreasing amplitudes:
\begin{eqnarray}
 C_{12-34} & =&G_1 G_3^* V_{13}+G_1 G_4^* V_{14}+G_2 G_3^* V_{23}+G_2G_4^*V_{24}\nonumber\\
&=&G_1 G_3^*+G_1 G_4^* +G_2 G_3^* +G_2G_4^* \nonumber\\
 && + G_1 G_3^* v_{13}+G_1 G_4^* v_{14}+G_2 G_3^* v_{23}+G_2G_4^*v_{24} \nonumber\\
&=&(G_1 +G_2)(G_3 +G_4)^* \nonumber \\
 && + G_1 G_3^* v_{13} - G_1 G_3^* v_{14} - G_1 G_3^* v_{23}+G_1G_3^*v_{24}\nonumber \\
 && +  G_1 G_3^* \epsilon_4^* v_{14} + \epsilon_2 G_1G_3^* v_{23} + \epsilon_2G_1G_4^*v_{24} + G_2G_3^*\epsilon_4^*v_{24} \nonumber\\
&=& G_1 G_3^* \epsilon_2\epsilon_4^*  \nonumber  \\
 && + G_1 G_3^* [v_{13} - v_{14} - v_{23}+v_{24} ] \nonumber \\
 && +  G_1 G_3^* [\epsilon_4^* v_{14} + \epsilon_2 v_{23} + (2\epsilon_2\epsilon_4^*-\epsilon_2-\epsilon_4^*) v_{24} ] 
\label{eq:C1234}
\end{eqnarray}
The first term of Eq.~(\ref{eq:C1234}) is the fully coherent part ($G_1 G_3^*\epsilon_2\epsilon_4^*$) of the electric field, corresponding to the flux from the main bright source which is leaking despite the null (eg, due to incorrect phasing). A loss of contrast due to the nature of
the incoming light, ie, $V\neq1$, is mainly encoded in the second term. It depends on $V_{1234}$ as:
\begin{equation}
V_{1234}=V_{13} - V_{14} - V_{23}+V_{24}=v_{13} - v_{14} - v_{23}+v_{24}\,. \label{eq:V1234}
\end{equation}
 The third and last term is different from zero if there is a combination of both  
a phase fluctuation and non-coherent light ($\propto \epsilon v$). It can be neglected with respect to the second term ($\propto v$) if the nulling is done at least partially : $\epsilon<<1$. It can be neglected with respect to the first term ($\propto \epsilon^2$) if the cophasing error is still dominant compared to the visibility loss due to the spatial coherency of the light: $v<<\epsilon$. This is the case when the null is limited by instrumental aberrations instead of the resolved astrophysical object. Something likely to happen when pushing the device to its limit for very faint companion observations.

Thus, neglecting this third term ($v<<\epsilon<<1$), we can approximate the triple product to:
\begin{eqnarray}
B_{12-34-56}&\approx& \left [G_1G_3^* (\epsilon_2\epsilon_4^*+V_{1234})\right]\times   \left [ G_3G_5^* (\epsilon_4\epsilon_6^*+V_{3456})\right]  \nonumber \\ 
             &&\times  \left [G_1G_5^* (\epsilon_2\epsilon_6^*+V_{1256})\right]^* 
\end{eqnarray}
Note that an additional simplification is possible if we assume small values for the closure phases ( $v << \epsilon^2$):
\begin{equation}
B_{12-34-56}\approx|G_1G_3G_5\epsilon_2\epsilon_4\epsilon_6|^2(1+\frac{V_{1234}}{\epsilon_2\epsilon_4^*}+\frac{V_{3456}}{\epsilon_4\epsilon_6^*}+\frac{V_{1256}^*}{\epsilon_6\epsilon_2^*})\,.
\end{equation}

However the phase of the triple product still depends on the unknown terms due to phasing errors ($\epsilon$). This is a real problem compared to classical closure phase which does not depend on optical aberrations. This is why it is necessary to diverge from the first concept presented in Fig.~\ref{fig:principle}: we have to cophase all pairs of telescopes. 

A way to do so is presented in Fig.~\ref{fig:Nul}. With respect to the previous scheme of Fig.\ref{fig:principle}, antenna 2 and antenna 3 are merged into one, as well as antenna 4 and 5. With that scheme, the whole array is phased (with $\pi$ delays): $G_2=G_1 (\epsilon_{2} -1)$, $G_3=G_2 (\epsilon_{3} -1)$ and $G_4=G_3 (\epsilon_{4} -1)$. From Eq.~(\ref{eq:C1234}), assuming $v << \epsilon << 1$ (but not necessarily $v << \epsilon^2$), the coherence terms can  be approximated to:
\begin{eqnarray}
C_{12-23}  &\approx& G_1 G_2^* \epsilon_2\epsilon_3^* + |G_0|^2 V_{1223} \\
C_{23-34}  &\approx& G_2 G_3^* \epsilon_3\epsilon_4^* + |G_0|^2 V_{2334} \\
C_{12-34}  &\approx& G_1 G_3^* \epsilon_2\epsilon_4^* + |G_0|^2 V_{1234} 
\end{eqnarray}
where $|G_0|$ is an average of the amplitude gain of all antennas. Hence, we can write the following relation:
\begin{eqnarray}
&( C_{12-23}-|G_0|^2 V_{1223}) ( C_{23-34}-|G_0|^2 V_{2334}) ( C_{12-34}-|G_0|^2 V_{1234})^*& \nonumber \\
&\approx& \nonumber \\
&|G_1G_2G_3\epsilon_2\epsilon_3\epsilon_4|^2&
\end{eqnarray}
or, keeping only the imaginary part:
\begin{equation}
 \Im \left[ \left( \cfrac{C_{12-23}}{|G_0|^2}- V_{1223}\right) \left( \cfrac{C_{23-34}}{|G_0|^2}- V_{2334}\right) \left( \cfrac{C_{12-34}}{|G_0|^2}- V_{1234}\right)^*\right]
\approx 0
\label{eq:ImRel}
\end{equation}

Thus, even if we cannot use directly the triple product (as well as the closure phase), this relation is a way to retreive the visibility information. 
According to it, the imaginary part of a combination of the coherence with the three terms $(V_{12}-1-V_{13}-V_{23})$, $(V_{23}-1-V_{24}+V_{34})$ and
$(V_{12}-V_{23}-V_{14}+V_{24})$ is always zero. These sum of visibilities are different
from zero in the case of a resolved object, and can be found, for example, by fitting in the least square sense the
$a$, $b$, and $c$ complex values into the relation:
\begin{equation}
 \Im \left[ (C_{12-23}-a) (C_{23-34}-b) (C_{12-34}-c)^* \right]=0 \,. \label{eq:Imabc}
\end{equation}
Note that, indeed, we are not using directly the triple product (as well as the closure phase). However, the spatial coherency information can be derived from this equation.

\subsection{Simulations}

   \begin{figure}
   \centering
   \includegraphics[width=5.6cm]{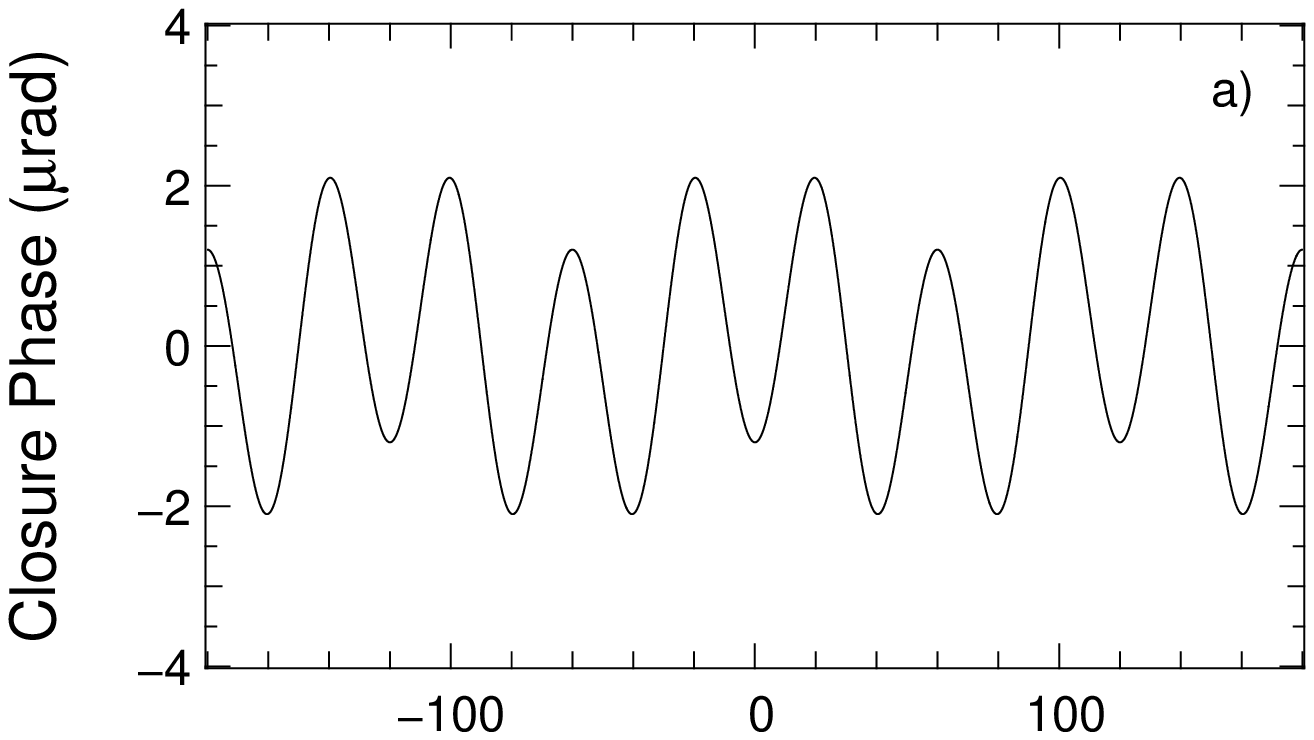} 
   \includegraphics[width=5.6cm]{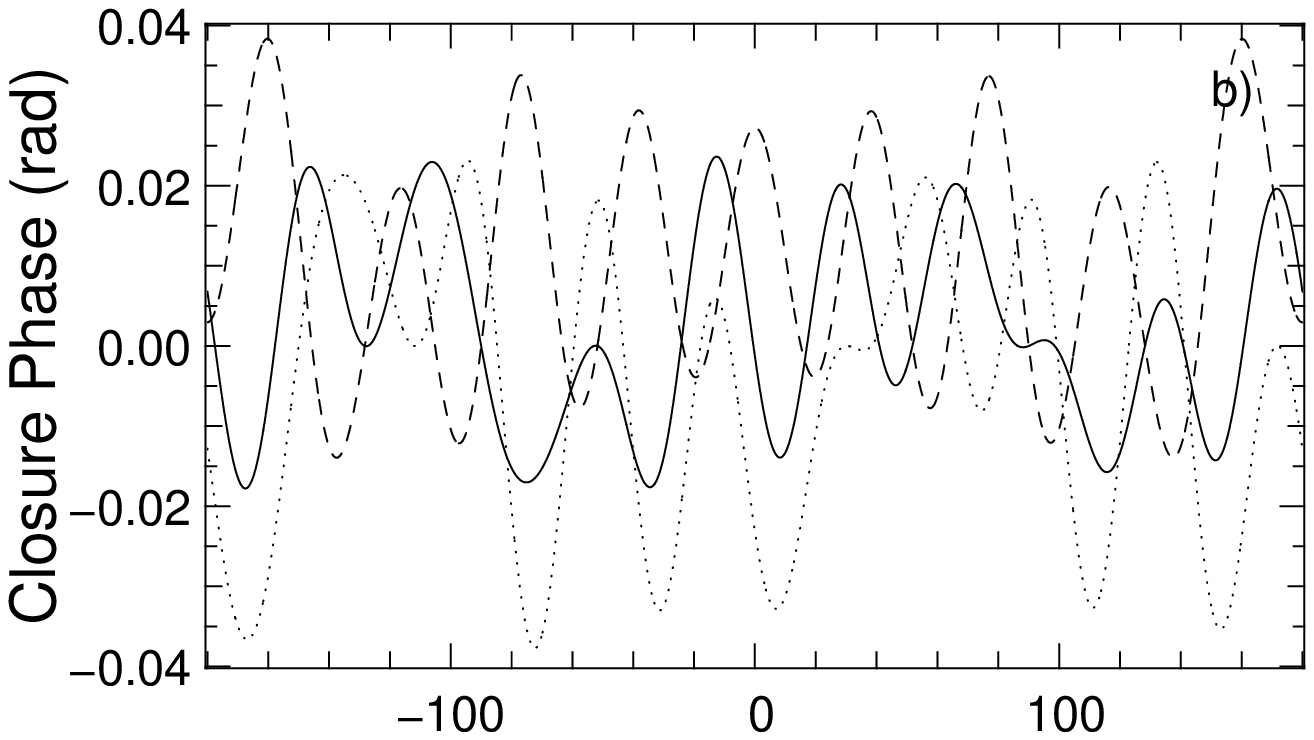}
   \includegraphics[width=5.6cm]{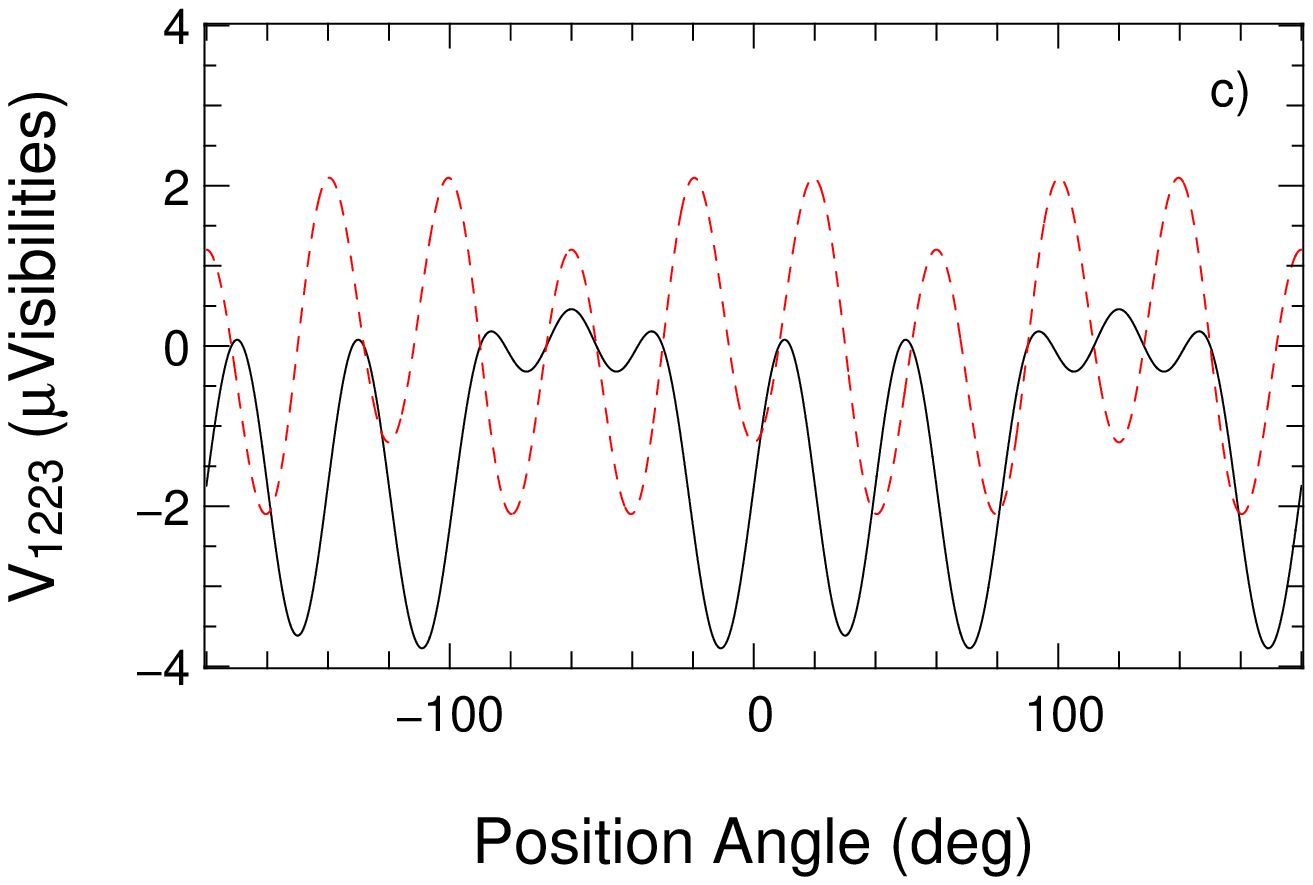}
      \caption{Stellar companion of flux $10^{-6}$ as would be observed by a triangle stellar interferometer of 1 meter baselines, as a function of the position angle of the companion. The three panels are, from top to bottom: a) the classical closure phases as calculated by Eq.~(\ref{eq:CPorig}). b) The closure phases on the nulled baselines as calculated by Eq.~(\ref{eq:tri}) and~(\ref{eq:CP}). The three curves correspond to 3 different cophasing errors of random values of the order of $10^{-2}$ the antenna gains. The amplitude of the closure phase is enhanced by a factor $\approx 10^{4}$. However, the signal is highly dependent on the perturbations.
c) The real (black) and imaginary part (dashed red) of the sum of visibilities as defined by Eq.~(\ref{eq:V1234}). Even if the amplitudes of the $V_{1223}$ is small (compared to the closure phase), the detection is easier thanks to the reduced photon noise on the nulled output.
}
         \label{fig:bin}
   \end{figure}

In this section, we compare the signal expected for a stellar companion on the i) classical closure phases, ii) baseline closure phases, and iii) sum of visibilities as obtained from Eq.~(\ref{eq:V1234}). To do so, we simulated a stellar interferometer with four entrance beams, as presented in Fig.~\ref{fig:Nul}. Projected toward the stellar target, the geometry of the interferometer appears as 3 baselines of 1 meter length rotated by $120\deg$ (corresponding to the baselines 12, 23 and 34 in Fig.~\ref{fig:Nul}). That is, beam 1 and 4 are situated at the same geographical location. The object is a star with a stellar companion $10^6$ times fainter, separated by 500mas at variable position angle. The wavelength is 1.55 $\mu$m.

The upper panel of Fig.~\ref{fig:bin} plots the classical closure phase as calculated by $\arg(C_{12}\cdot C_{23}\cdot C_{31})$. Panels b) and c) of the same figure show the baseline closure phase $\arg(C_{12-23}\cdot C_{23-34}\cdot C_{12-34}^*)$ and the visibilities $(V_{12} - 1-V_{13} +V_{23})$ (the real part in black, and the imaginary in red). Each plot is a function of the position angle of the target relative to the orientation of the interferometer (perpendicular to baseline 23). The upper and lower panels do not depend on the level of nulling. The small scales ($\mu$radians for the first plot, $10^{-6}$ in visibility amplitude for the second) are due to the faintness of the companion. On the other hand, the closure phases as calculated in the middle panel are much higher. They depends heavily on the level of the nul. For example, in solid black are the coherence term thanks to Eq.~(\ref{eq:C1234}) using a level of nulling of the order of $10^2$: $\epsilon_2=0.014$ and $\epsilon_3=-0.013i$ and $\epsilon_4=0.02$. The dashed and dotted lines correspond to other $\epsilon$ values of similar amplitudes.

The closure phase signal is stronger in the middle panel with respect to the upper panel. It is a consequence of the nulling stage. By using a nulling at the level of $10^2$ on the gain of each antenna, the signal increase by a factor $\approx10^4$. However, as discussed in the previous section, the nulled closure phase are such that they depend heavily on the error in phase and amplitude. This explains why we are using the third panel, the visibilities, that we can adjust in Eq.~(\ref{eq:Imabc}). Even though that the $V_{1223}$ are of the micro level amplitude, their detection is easier because the spatial coherency values $C_{1223}$ are also very low thanks to the nulling stage. With respect to the effect of photon noise, measuring $V_{1223}$  from the $C_{1223}$  values is therefore easier than  measuring the closure phase from the $C_{12}$ values.

\section{A practical implementation}   

\subsection{The Nuller Stage}
\label{sc:nul}

   \begin{figure}
   \centering
   \includegraphics[height=4cm]{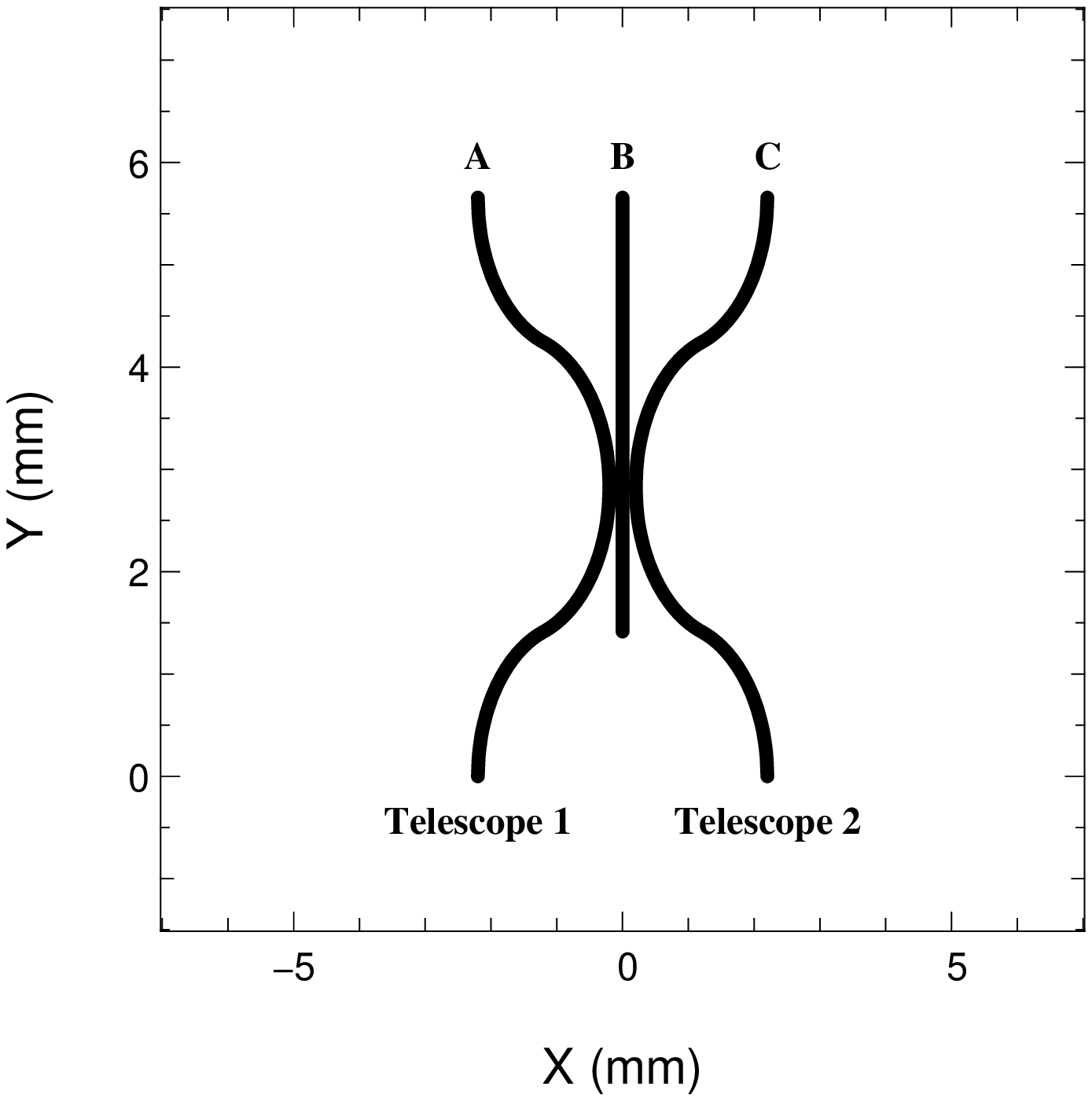} \hspace{0.8cm}
   \includegraphics[height=4cm]{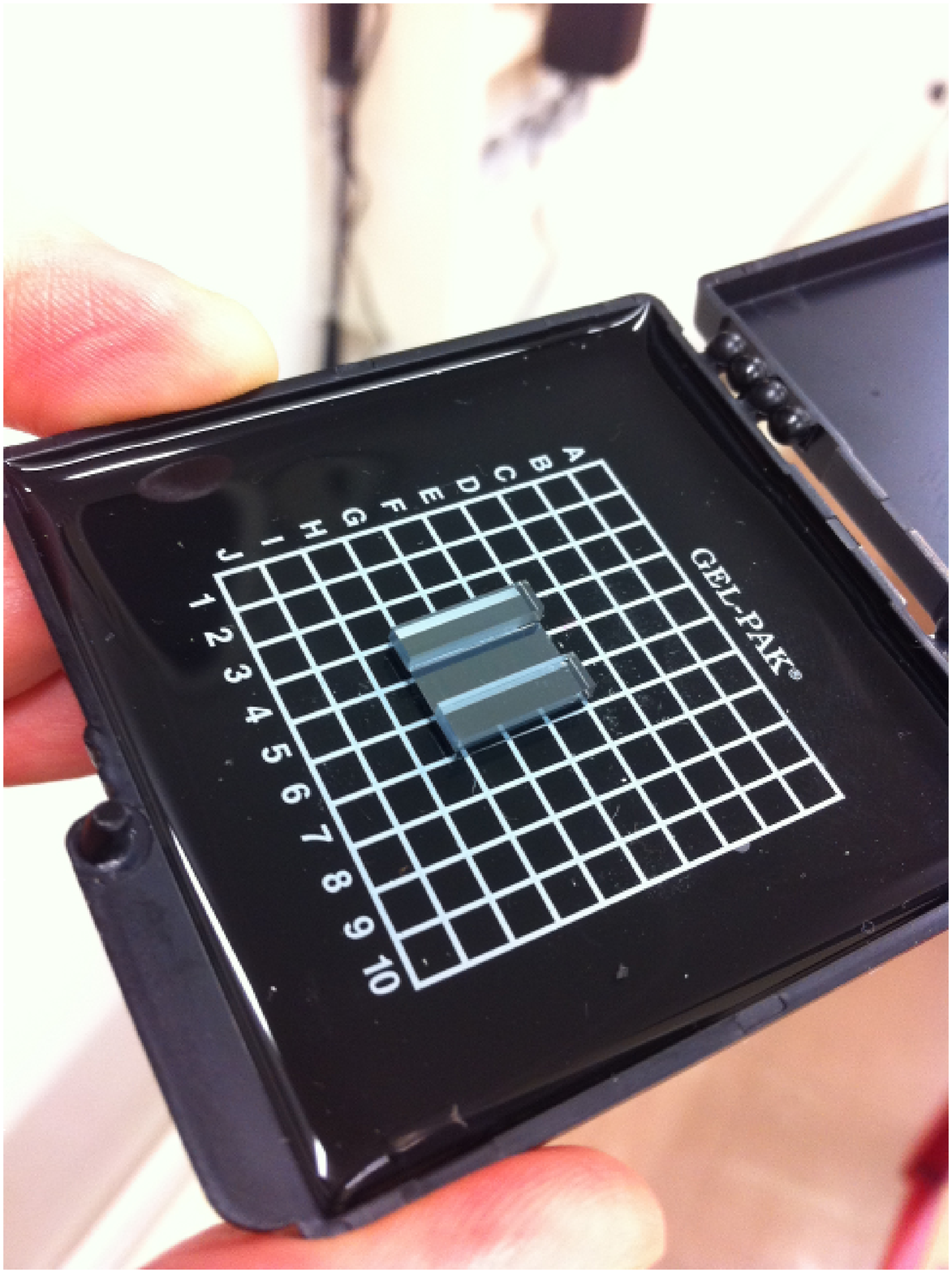} \\
   \includegraphics[height=4cm]{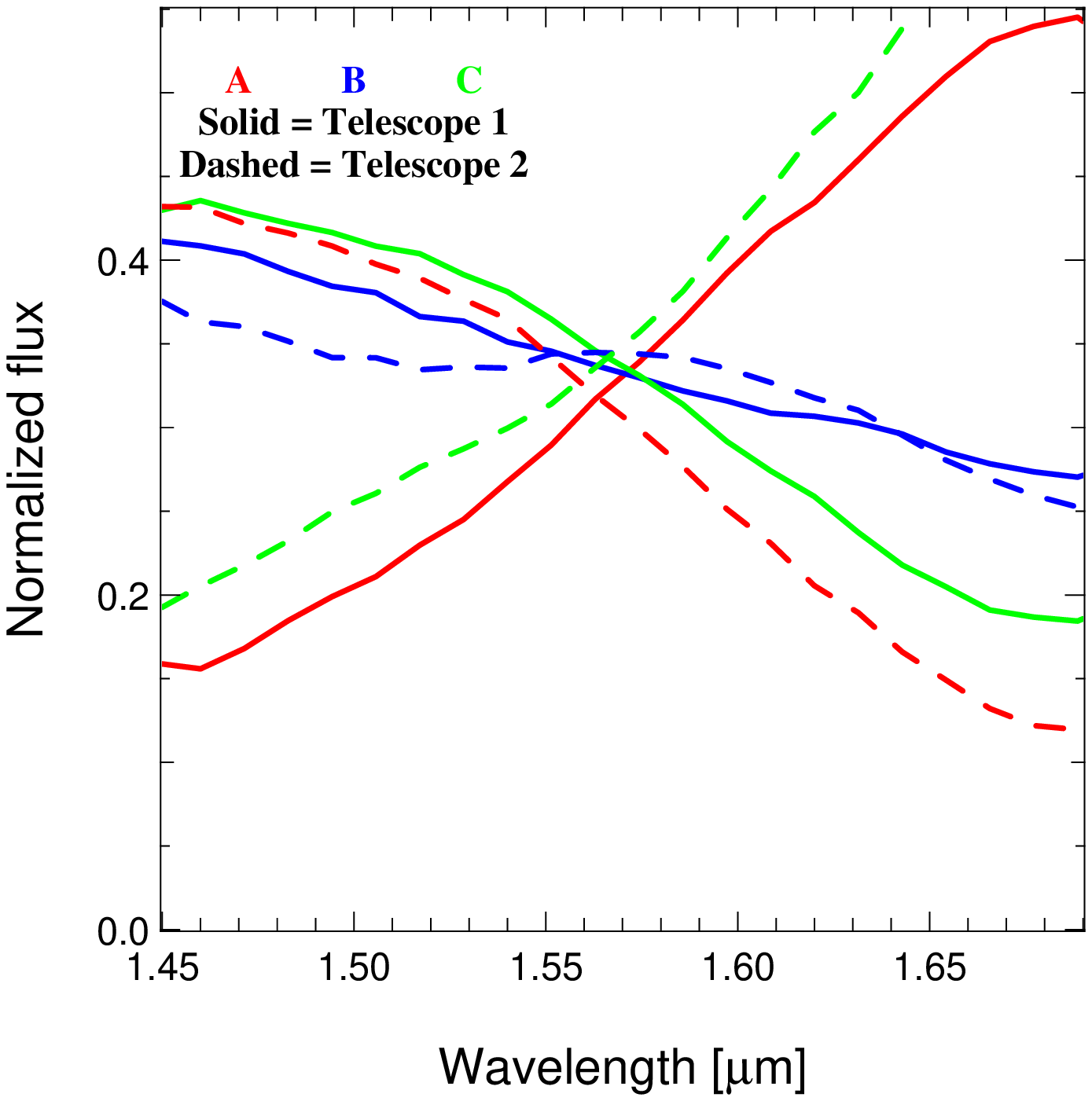}
   \includegraphics[height=4cm]{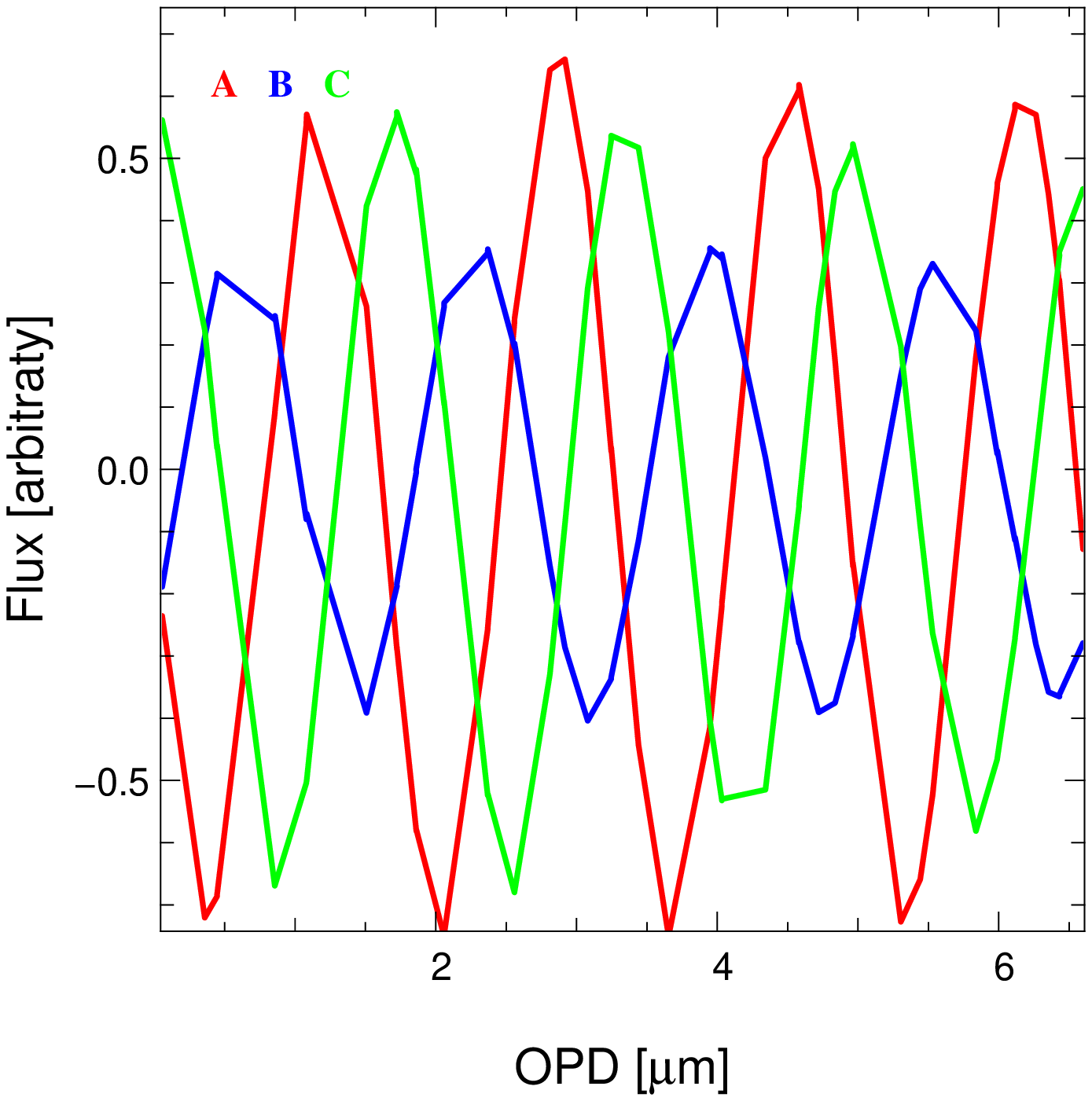}
      \caption{Upper left panel: routing of a tri-coupler. The light coming from the two telescope entrances 1 \& 2 is
coupled by evanescence into the three outputs \textbf{A}, \textbf{B}, and \textbf{C}. Upper right panel: photo of the
integrated tricoupler tested in this paper. Lower left panel: normalized flux on the three
outputs of the ABC beam combiner. An ideal ABC beam combiner would split flux equally into the three outputs, as
happens here at 1.56 $\mu$m. At this wavelength the phase
shifts between A and B and C and B are 120 degrees (lower right panel). }
         \label{Tri}
   \end{figure}

In the following sections of this paper, we explore the performance of such an instrument if based on the integrated optics 
(IO) technology. This is not a requirement for the concept presented in Section~\ref{sc:concept}, but the combination of
single mode filtering and extreme stability makes this technology an excellent vehicle for this concept. 
The basic idea is that starlight can be injected into an array of polarization-preserving single mode fibres, 
as was done in the FIRST \citep{2012A&A...541A..55H} and Dragonfly \citep{2012MNRAS.427..806J} instruments. 
The light can then be described by a limited number of
properties (the amplitude, the wavelength, the phase, and the polarization vectors). As in any astronomical
interferometer, the value of importance is then the complex coherence values between each of the beams, since they
convey the spatial information of the astronomical object.
A classical optical interferometer would simply combine the beams and extract visibilities (the coherence values).
However, as explained in Section~\ref{sc:concept}, we also incorporate a first interferometric stage which extincts on-axis starlight before
combining the beams to extract visibilities. This first nulling stage is based on a tri-coupler. 

A tri-coupler was first proposed for beam combination by \citet{2009arXiv0904.3030L} \footnote{available online: http://arxiv.org/abs/0904.3030}. 
It was subsequently proposed for nulling by \citet{2010ApOpt..49.6675H}. 
The routing is presented in the upper left panel of Fig.~\ref{Tri}.
It works by evanescent coupling of light between one guide into three output waveguides.
The lower left panel of the same figure shows the measurements obtained on a test device made by NTT electronics. 
The IO device was calculated to have a 33.3/33.3/33.3\%\footnote{the coupling ratios are given for non-interfering light, ie when feeding only one arm of the tricoupler.} beam splitting ratio at a wavelength of 1.55 $\mu$m. The idea is to use the A and C
outputs for fringe tracking (where almost all the flux will be present) while on the B output the flux is extinguished
by coherent combination. A $\pi$ phase shifter can be either included inside the IO (and hence opening the possibility
for a design which is achromatic), or outside the IO by the means of the delay lines already necessary for phase
tracking (this is the option chosen here).

An important advantage of the tricoupler is the three outputs: one null and two bright outputs give a perfect 
configuration for phase sensing. At 1.55 $\mu$m, the phase shifters between A\,\&\,B and C\,\&\,B are around 120 degrees, with the flux equally split between the 3 outputs.
This is an ideal compromise between two extremes: a 50/0/50\% or a 25/50/25\% beam combiner. On one hand the 50/0/50\% would have phase
shifters equals to -90/0/90 degrees: perfect for fringe tracking but bad for nulling since no flux would be conveyed
into the B output. On the other hand, the 25/50/25\% would maximize the output into the B values, but the energy
conservations principle would force the phase shift to -180/0/180 degrees, hence making it useless for fringe tracking.

\subsection{The Closure Phase Stage}
\label{sc:bc}

   \begin{figure}
   \centering
   \includegraphics[width=6.cm]{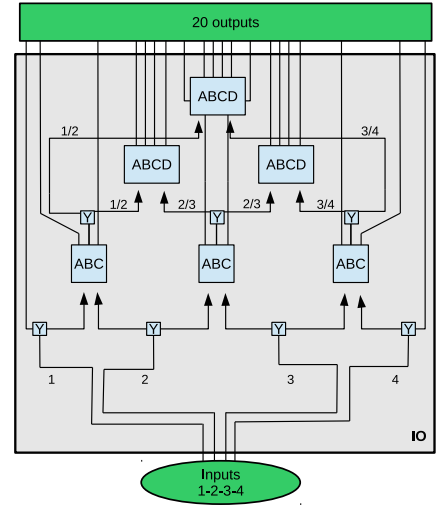}
      \caption{
IO version of the principle shown in Fig.~\ref{fig:Nul}: inputs 2 and 3 are split into two to make six beams input to the nuller 
stage as before. 
Beams 1 and 4 are also split at the same first layer of Y junctions, but only one of the two output channels proceeds to the 
nuller for this case.
For this configuration, the nuller stage interferes beams 1--2, beams 2--3 and beams 3--4 in the ABC tri-couplers.
The final stage of the integrated optics accepts the 3 nulled ``B'' beams from the tri-couplers, which are again split
in two, and obtains the complex coherences (and therefore closure phases).
This beam recombination is done with the well-known ABCD beam combiner to produce the coherency values 
$C_{12-23}$,  $C_{12-34}$ and $C_{23-34}$.}
         \label{fig:IO}
   \end{figure}

The second part of the optical setup is geared to generate closure phase data. The basic idea is that due to
phase errors, the nulls will not be perfect, and the best way to test the light for spatial structure is to generate
closure phases: in the case of a perfectly coherent source, the closure phase will be zero. In the presence of an asymmetry
(such as a high contrast companion), the closure phase will in general produce a non-zero signal (see demonstration
Section~\ref{sc:concept}).
 
To derive closure phases, at least three beams are necessary. Since we want to have a nulling stage before, it means
that 6 inputs are in theory necessary, as shown in Fig.~\ref{fig:principle}. However, as demonstrated in Section~\ref{sc:triple}, it is necessary to cophase all the telescopes to have a meaningful observable. Thus, we are using a schematic optical concept with only 4 input beams, as presented in Fig.~\ref{fig:Nul}.
 Practically, each beam is divided into two, before making interference to get
the nulls between pairs of telescopes. So beam 2 is combined with 1 \& 3, and beam 3 with 2 \& 4. To respect the equilibrium in
flux (required for a good null), beam 1 and 4 are also split in 2, but half on the light is only used for photometric 
measurement.  Each null output of the pairwise combinations are then combined with the others to derive the complex visibilities
which will be used to compute the closure phases. On an IO device, it appears as shown in Fig.\ref{fig:IO}. 
 
\begin{figure}
   \centering
   \includegraphics[width=6.5cm]{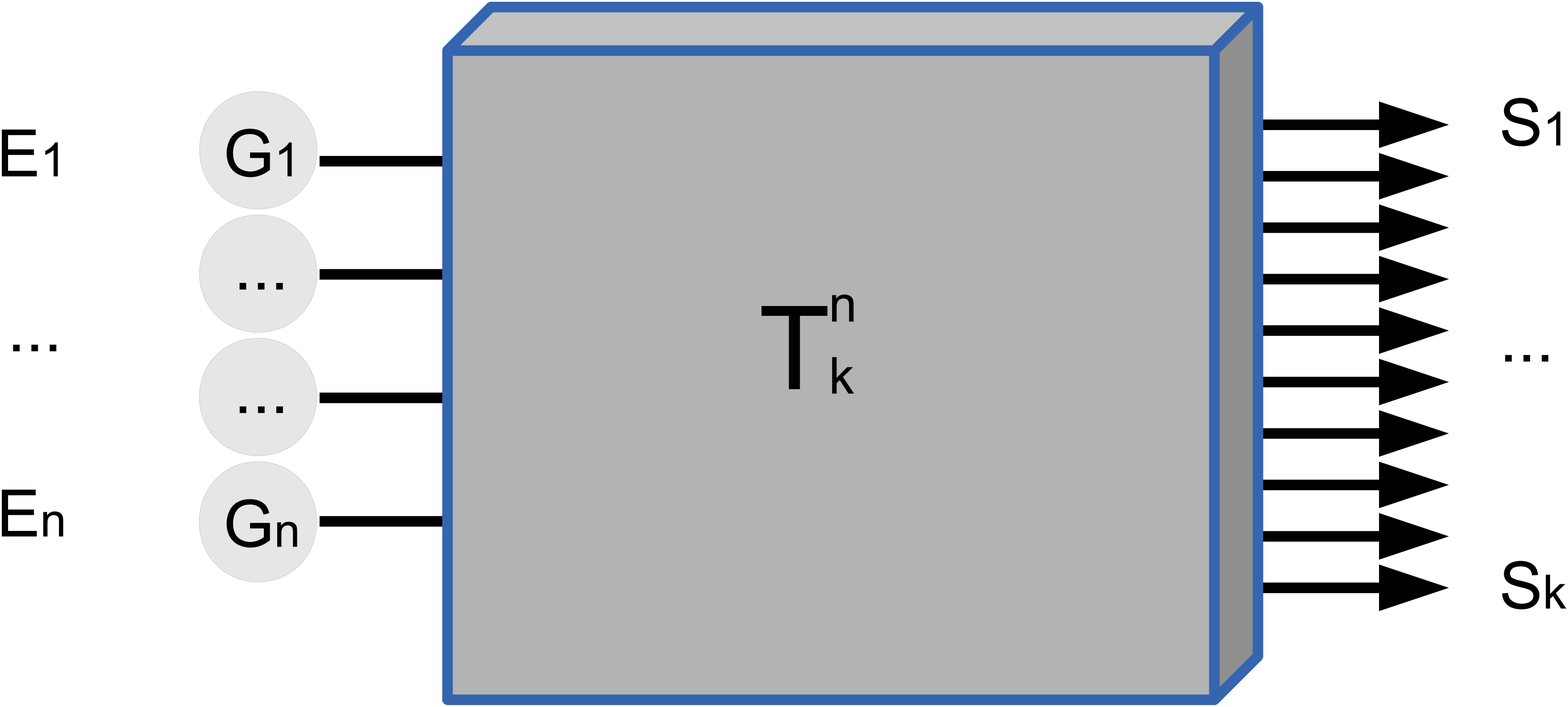}
      \caption{Theoretical representation of the optical setup. The relationship between the electric field at the entrance and the electric field at the output is $S_k=\sum_n T_k^n G_n E_n$. The complex antenna gains are the $G_n$. The optical transfer function of the instrument is encoded into the $T_k^n$ values. }
         \label{fig:simpIO}
   \end{figure}

To model the effect of this complex beam combiner on the stellar light, we followed the nomenclature from
\citet{2008SPIE.7013E..31L}. To be coherent with Section~\ref{sc:basic}, we updated the principle to account for the antenna gains. Hence, the relationship between the electric field at the entrance of the interferometer ($E_n$) and the electric field at the output ($S_k$) is now:
\begin{equation}
S_k=\sum_n T_k^n G_n E_n
\end{equation}
where both the transfer function $T_k^n$ and the antenna gain $G_n$ are represented in Fig.~\ref{fig:simpIO}. On the detector, we have a measurement of the amplitude of the electric field, average over the detector integration time:
\begin{equation}
<|S_k|^2>=<|\sum_n T_k^n G_n E_n|^2>
\end{equation}
where $<>$ represent an average value over the detector integration time: long with respect to the coherence time of the electric field ($<~E_i~>\approx 0$), but short with respect to variations of the gains ($<~G_i~>\approx G_i$). This relation can be expanded to obtain a linear expression:
\begin{equation}
<|S_k|^2>=\Re \left[ \sum_n |T_k^n|^2 | G_n|^2 <|E_n|^2> + 2 \sum_n\sum_{m>n} T_k^n T_k^{m*} C_{nm} \right]
\end{equation}
where   $C_{nm}$ are the
coherence terms defined in Eq.~(\ref{eq:C12a}).

Similarly, there is a linear relationship between the flux observed on the 20
outputs of the IO of Fig.~\ref{fig:IO} and the 4 electrical input and antenna gains. It can be written as a product of two matrices:
\begin{eqnarray}
\left(
\begin {array}{c}
<|S^1|^2>\\
\vdots \\
<|S^K|^2>
\end {array}
\right)
&=  &
\Re \left[ 
V2PM_a
\cdot
\left(
\begin{array}{c}
|G_1|^2<|E_1|^2>\\
|G_2|^2<|E_2|^2>\\
|G_3|^2<|E_3|^2>\\
|G_4|^2<|E_4|^2>\\
C_{12}\\
C_{13}\\
C_{14}\\
C_{23}\\
C_{24}\\
C_{34}\\
\end{array}
\right)
\right]
\label{V2PMEq} 
\end{eqnarray}
where the $V2PM_a$ is a 20 by 10 matrix made of a combination of the complex values $T_k^n$. The term $V2PM$ stands for "Visibility to Pixel Matrix", and is a name frequently used in interferometry to address the problematic of the transfer function of an interferometer \citep[eg,][]{2007A&A...464...55T}.

Alternatively, the $C_{ij}$ terms can be combined into baseline coherence terms $C_{ij-kl}$ using the linear relation spelled in Eq.~(\ref{eq:C1234decomp}). The $V2PM$ matrix is then changed according to the following equation:
\begin{eqnarray}
\left(
\begin {array}{c}
<|S^1|^2>\\
\vdots \\
<|S^K|^2>
\end {array}
\right)
&=  &
\Re \left[ 
V2PM_b
\cdot
\left(
\begin{array}{c}
|G_1|^2<|E_1|^2>\\
|G_2|^2<|E_2|^2>\\
|G_3|^2<|E_3|^2>\\
|G_4|^2<|E_4|^2>\\
C_{12}\\
C_{23}\\
C_{34}\\
C_{12-23}\\
C_{23-34}\\
C_{12-34}
\end{array}
\right)
\right]
\label{V2PMEb} \,.
\end{eqnarray}
Note that the $V2PM_b$ is not anymore the {\em visibility to pixel matrix} in the strict sense. Both $V2PM_a$ and $V2PM_b$ are given for the beam combiner presented Fig.~\ref{fig:IO} in the Appendix. Of these two relations, the complex $V2PM_a$ matrix is equivalent to the optical transfer function of the beam
combiner. It depends on the routing of the beam combiner, and can be calculated from theory according to the mapping
presented in Figure~\ref{fig:IO}. Its terms are explicitly evaluated in Eq.~\ref{eq:V2PMa}.

The $V2PM_a$ is complex matrix, and can be separated into a real and imaginary part. It then becomes a 20 by 16
matrix. This matrix is over-constrained, and as
such, is uniquely invertible in the least squares sense (by a singular value decomposition for example). However, inversion
of this matrix does not preserve the fact that photon noise is input specific, and each ``ABCD'' has a smaller level
photon noise thanks to the low number of photons at these outputs. If we want to extract the $C_{ij-kl}$ information
separately, we have to use the $V2PM_b$ which is a linear combination of $V2PM_a$. 

\subsection{Dynamic range}
\label{sc:res}

   \begin{figure}
   \centering
   \includegraphics[height=4cm]{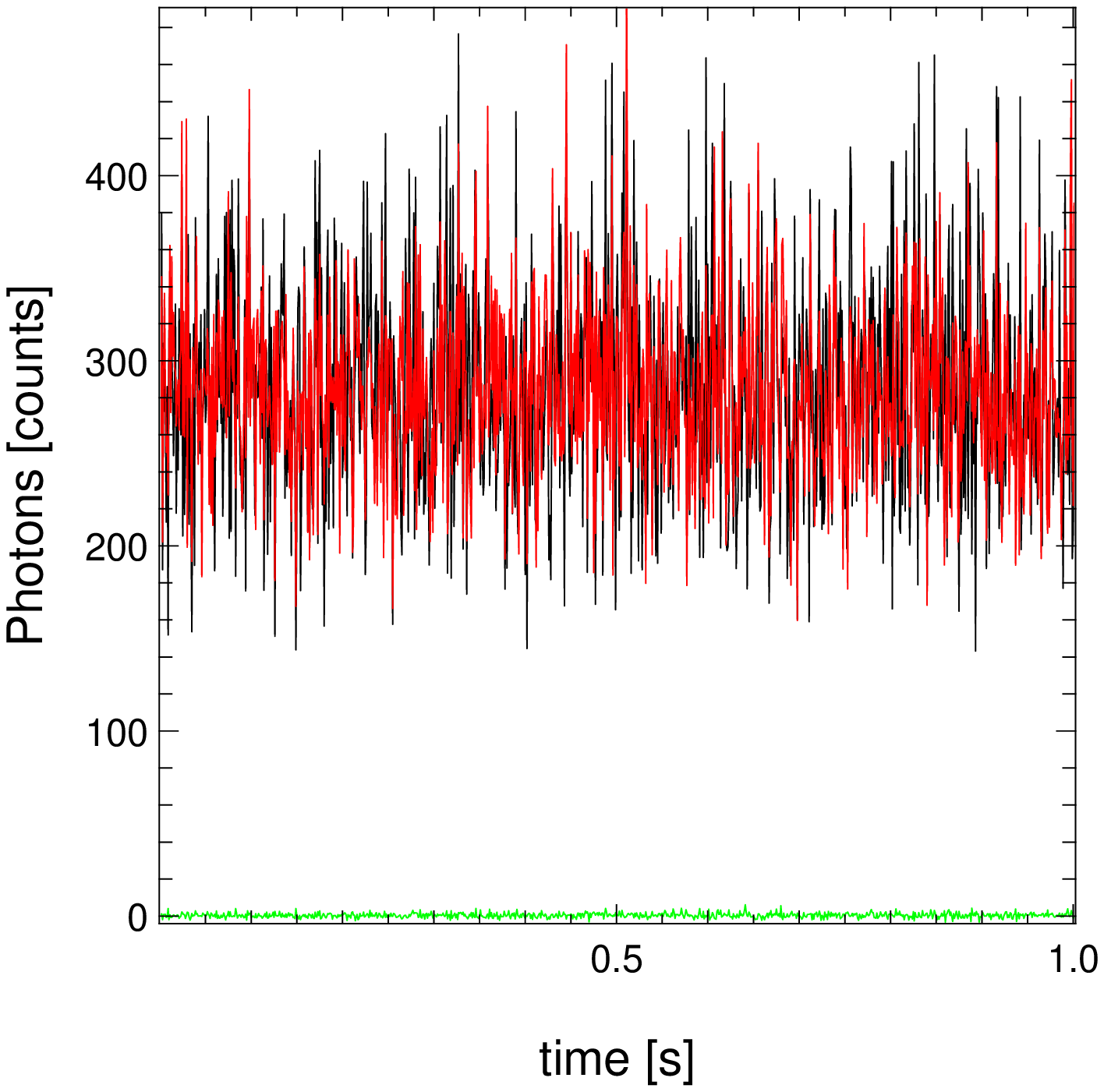}
   \includegraphics[height=4cm]{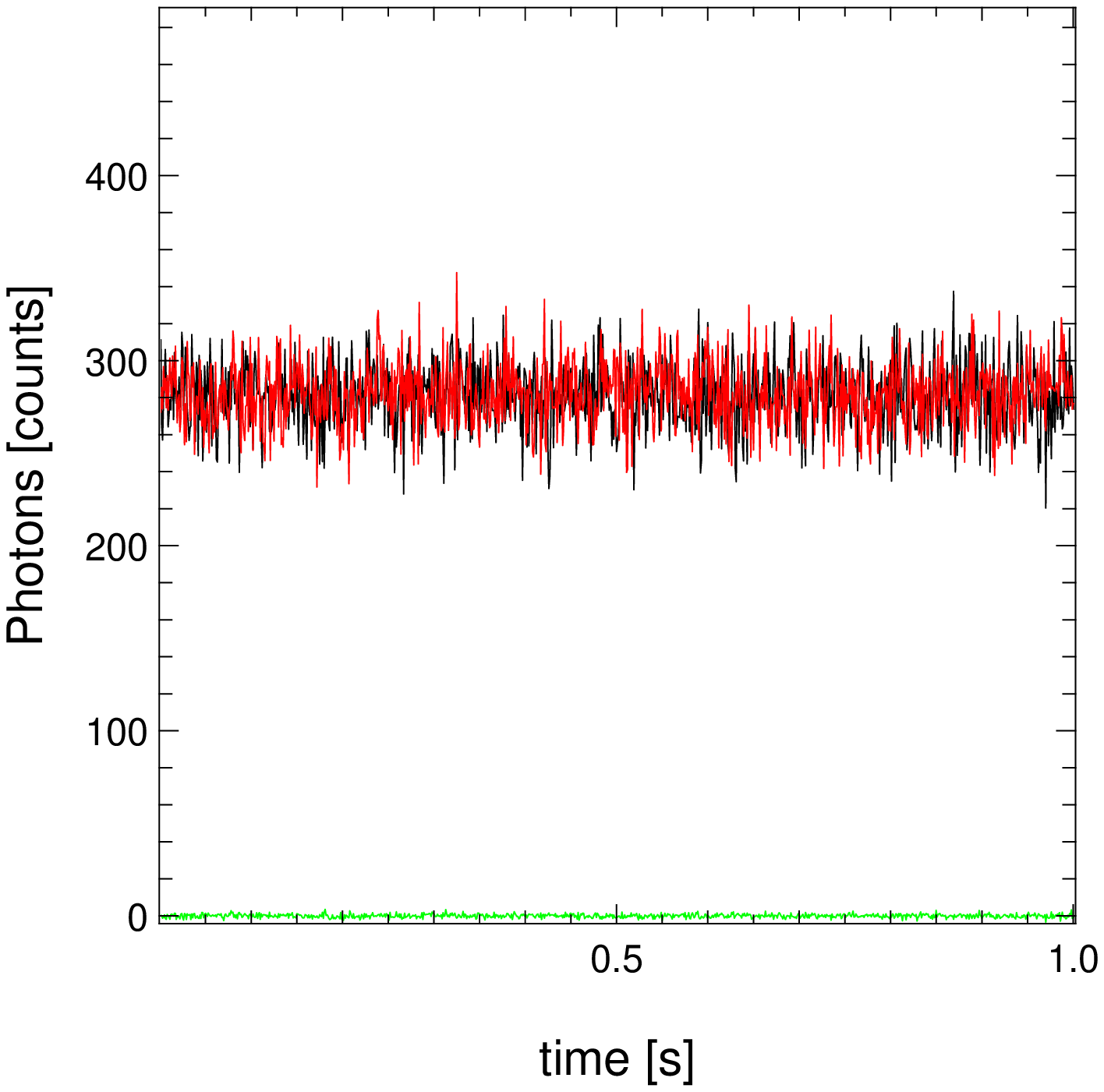}
      \caption{Simulated data illustrating the theoretical performance of our beam combining architecture.
Data depicts outputs from channels 1, 3 and 5 (respectively black, red and green) from the beam combiner which correspond 
to the first, third, and fifth line of the $V2PM_a$. These are respectively: i) the photometric channel of input 1, 
ii) the A output of the first ABC beam combiner, and iii) the second output of the first ABCD. 
Output signals were calculated using the $V2PM_a$ matrix with input the electric field of following properties. 
Absolute phase in each exposure was random, but differential phase between 1 and 2, between 2 and
3, et between 3 and 4: value $\pi$ with 0.1 rad error (left panel) and 0.01 rad error (right panel). 
The average flux ratio between the null outputs (green) and the
A output (red) is 100 (left panel) and 10\,000 (right panel). }
         \label{Photons}
   \end{figure}

We performed numerical simulations of the beam combiner architecture presented in the previous section. For
comparison, we also modelled a classical 4 beam combiner as used by the GRAVITY interferometer
\citep{2011Msngr.143...16E,2012SPIE.8445E..2XJ}. Our
beam combiner is modelled using the matrix $V2PM_a$. The ``classical'' 4-beam combiner is modelled by the $V2PM_{\rm
GRAVITY}$ in Eq.~(\ref{GRAVITY}) of the Appendix. It is the same one stated in Eq.~(6) of \citet{2008SPIE.7013E..31L}.

The model consists in a point-like target ($C_{ij}=G_i G_j^*$). Phasing errors due to atmospheric aberration or instrumental effects are simulated using
an arrays of 10\,000 $G_1$, $G_2$, $G_3$ and $G_4$ complex antenna gains. 
A total of 25$\times$8 of such arrays were computed. First, different conditions of brightness of the source were
simulated by 25 values ranging from 10 to 10$^8\ $e$^-$ (detected photons) per integration time. Second, different
level of phase tracking and injection stability were simulated by 8 regimes of errors ranging from 0.2 to 0.001
radians rms (accompanied by an error ranging from 0.2 to 0.001 rms in amplitude ratio). From these arrays, signals at
the output of the integrated optics were computed from Equation~(\ref{V2PMEq}). 
The same input arrays were used to compute two datasets: for the first, we used $V2PM_a$, and for the second we used 
$V2PM_{\rm GRAVITY}$. On each dataset, photon noise was added, plus a background noise taken as $\sigma_{\rm readout}=1
$e$^-$ per sample. A subset of such dataset is plotted in fig.~\ref{Photons}. It shows three outputs of the  $V2PM_a$ for two different tracking accuracies.

\begin{figure}
   \centering
   \includegraphics[height=4.05cm]{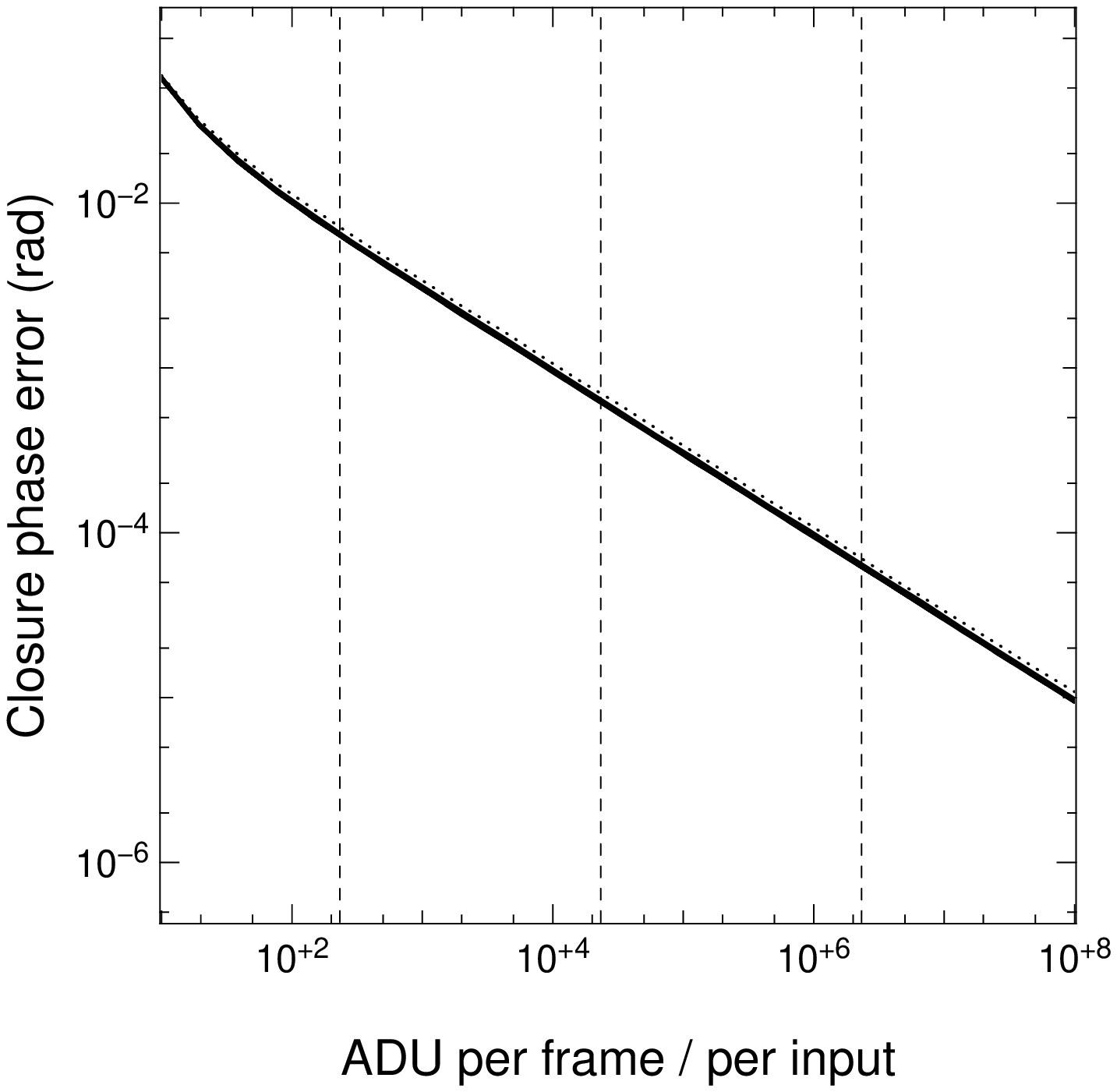}
   \includegraphics[height=4.05cm]{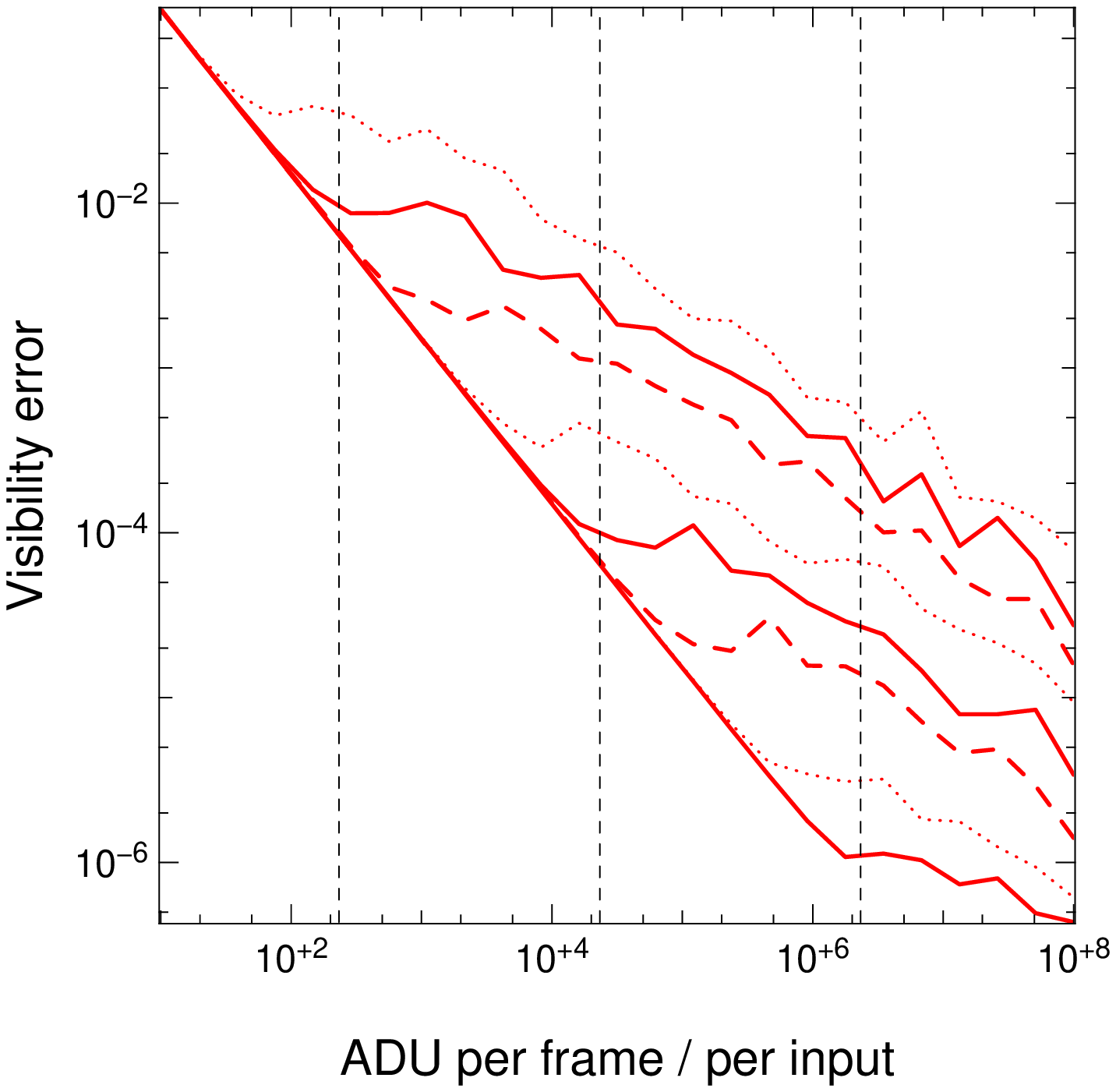}
      \caption{Accuracy obtained on the closure phase (upper panel, on a "GRAVITY-like" classical 4 beam combiner) and visibilities (bottom panel, on the beam combiner described in Fig.~\ref{fig:IO}) as a function of the number of ADU per frame. 
The different curves correspond to different level of phasing. On a classical beam combiner, the closure phase eliminate the piston error, hence there is no dependency on the accuracy of the fringe tracking. However, for a beam combiner including a nuller, the phasing is important. The phasing
error for each red curve from the top down corresponds to 0.2 (dot), 0.1 (solid), 0.05 (dash), 0.02 (dot), 0.01 (solid), 
0.005 (dash), 0.002 (dot), 0.001 (solid) radians. The closure phase slope is a function of the square of the photon number. The red curves have two regimes: a
read-out noise regime, where the slope is a linear function of the number of photons. And a photon noise regime, 
where the limit comes from the remaining photon noise after nulling (the slope is a linear function of the square 
root of the number of photons).
The better the phasing the lower the transition between the two regimes.}
         \label{Result}
   \end{figure}

From the second dataset (simulating the GRAVITY beam combiner), we computed the closure phases in a ``classical'' way. 
To do so, we multiplied the output flux with the inverse matrix of the $V2PM_{\rm GRAVITY}$ matrix. Then, we get 6
arrays of complex visibility values, which are multiplied to give 4 triple products (the bispectrum). These triple
products are then summed over a batch of 1000 integration times, and the closure phase is extracted by getting the
arguments. Plotted in black in Fig.~\ref{Result} is the accuracy of the closure phase measurement (in radians)
calculated by taking the rms value of the different batches. As expected, the accuracy of the closure phase does not
depend on the phasing of the inputs. Except at very low flux (Nphoton $< 10$ per frame), the accuracy follows a trend
proportional to the square root of the number photons. This is the photon noise regime \citep[an analytical description
of photon noise can be found in][]{2013MNRAS.433.1718I}.

From the first dataset (simulating our new beam combiner using $V2PM_a$), we used the inverse matrix of $V2PM_b$ to
derive the complex value of $C_{12-23}$, $C_{23-34}$ and $C_{12-34}$. We then used a Levenberg-Marquardt algorithm to
obtain the $a$, $b$ and $c$ values which minimize the $\chi^2$:
\begin{equation}
\chi^2=\sum_1^{1000}{\left[
\Im [ (C_{12-23}-a) ( C_{23-34}-b) (C_{12-34}-c)^*]
\right]^2} 
\label{chi}
\end{equation}
Finally, the $a$, $b$ and $c$ values are normalized by the average transmission $|G_0|^2$, and the error obtained on the
complex value is compared with the CP error in Fig.~\ref{Result} (red curves). 

Unsurprisingly, the dynamic range depends on the quality of the phase tracking. If the phase tracking is good, the
accuracy -- and hence the dynamic range -- increases linearly as a function of the photon number. This is because we are
in the readout limit regime: the dynamic range is proportional to the flux of the source. In Fig.~\ref{Result} this
limit comes from the 1$e^-$ readout noise and comes into play when the amplitudes of $C_{12-23}$, $C_{23-34}$ and
$C_{12-34}$ are of that order. However the phase tracking is sometimes not enough to reach that floor. Under these
conditions, the accuracy of the $a$, $b$ and $c$ values are photon noise limited and decrease as a function of the
square root of the number of photons -- as happens with classical closure phases.

An interesting point to note is where the black curve cross the red curve, ie, when nulling becomes more efficient than
classical closure phase calculations. This point depends on the setup, and the configuration of the integrated optic 
(notably the tri-coupler equilibrium). However, it can be empirically estimated for a general setup. To do so, we have to consider
that 2 thirds of the light is lost for fringe tracking -- this is equivalent to an increase of a factor 9 in the readout
noise. For that loss to be compensated by the gain in terms of photon noise, the extinction must be a factor 9$^2 \times
\sigma_{\rm readout}$ on the nulled outputs. Hence, it means that the setup must receive at least 81\,e$^-$ per
integration time. This is roughly what we observe in Fig.~\ref{Result}.

\section{Discussion}
\label{sc:con}

\subsection{Applications}

Our beam combiner setup makes no difference between a multi-telescope long baseline interferometer, or a fibred pupil
remapper as proposed by \citet{2006MNRAS.373..747P} or \citet{2002RSPTA.360..969B}. Hence, it can be used for both
instruments. However, as mentioned in Section~\ref{sc:res}, for the nulling to be interesting, the gain in terms of
photon noise must outweigh the loss due the photons discarded for the fringe tracking. This tradeoff can be
empirically determined. It roughly corresponds to a regime where the number of photons detected per integration time and
per aperture is equal to 100 times the detector readout noise.

This means that this kind of interferometric nulling is valid either for long integration times, bright sources, or low
readout noise detectors. Space applications may present a logical venue for this technique. Instead of modulating a 
nulled signal by rotating the interferometer, the closure relation would instantaneously probes the coherence of the 
light, and discriminates whether the residual flux beyond the nuller stage is due to bad phase tracking or spatially 
incoherent light due to the presence of a companion. Thanks to the relative stability of optical wavefronts in space,
the integration times could be long enough to fully benefit from the nulling
stage.

A second application is envisaged from ground-based visible wavelength observations. Thanks to low noise L3CCD type camera, 
readout noise can be low enough even with respect to the few photons collected during the integration time. However,
the drawback of observing at visible wavelengths is that the coherence times are small (hence short integration time) and
the cophasing difficult.

Lastly, an interesting development of this technique could be at near-infrared wavelengths. The new
generation of IR avalanche photodiodes detectors \citep{2012SPIE.8453E..0TF} has shown readout noises below 1\,e$^-$.
In the near-IR the atmospheric phase errors are more benign, and integration times can be longer. For example, in 10\,ms,
 a 10 magnitude star delivers more than 2000 photons into a one meter square aperture in the K band. So to apply our
rule of thumb, nulling could be useful on these kind of targets once the detector readout noise falls below 5\,e$^-$. 
A more complete analysis would need to include throughput, spectral dispersion, quantum efficiency, and other technical
details.

\subsection{Possible improvements}

Several improvements of this scheme can be imagined. The first is a more complex setup. Fig.~\ref{fig:IO} presents the
simplest version of the concept with only 4 inputs, but we can imagine a similar combiner with any number of beams.
Only the complexity of the routing would increase, and the size of the $V2PM$. Concerning the integrated optics device, 
another interesting horizon of research is active optics (as already offered by Lithium Niobate), offering on-the-chip amplitude and phase control.

The second possible enhancement is to apply a weighting when fitting the $a$, $b$ and $c$ complex values
in Eq.~(\ref{chi}). To be more accurate, it is also possible to decompose the complex values into a sum of real
values (eg., $a=a_r+j.a_i$) and to take the imaginary part of $(C_{12-23}-a) (C_{23-34}-b) (C_{12-34}-c)^*$ to
get to a classical non-linearity optimization problem. Once generated, such weightings could then be used to 
preferentially emphasize integrations where the cophasing is at its best.

Finally, a possible arithmetical trick could be to stick with the description of the $V2PM_a$, to determine
the six complex values $G_{i}G_{j}^*V_{ij}$, and use the covariance matrix to project the data on a space of minimal
variance. This would gives the best dynamic range whenever the phasing is good or the photon number low. This is
under study, and may be the subject of future work. But even if more elegant, this will not change the
general conclusions of the present paper.

\subsection{Conclusions}
   
   In this paper we presented a setup for a nuller which can be described by two optical stages: 
   \begin{enumerate}
    \item A nulling stage, with the goal of extinguishing axial light of the central bright star.
    \item A set of ABCD beam combiners which combine the nulled outputs of the preceding stage, with the goal of 
characterizing the coherence of the remaining light in a manner robust against imperfect cophasing of the incoming 
stellar light.
   \end{enumerate}

We showed that such a system would have the advantage of low photon noise, and no phase noise. Instead of 
using the triple product of the bispectrum, we propose to fit the relation
\begin{equation}
 \Im [ (C_{12-23}-a) ( C_{23-34}-b)(C_{12-34}-c)^*]=0
\end{equation} to the data.
It relies on the coherence terms between the nulled output (the $C_{ij-kl}$), and it has properties similar to closure phases. The $a$, $b$ and $c$ values can then be directly linked to the object visibilities, or more precisely to a combination of four visibilities. For example, $a$ is proportional to $ V_{13}-V_{23}-V_{14}+V_{24}$.

This setup could be interesting for ground based high contrast direct detection of companions around bright stars, 
and also has potential as a space interferometer architecture.

\section*{Acknowledgments}
This work was supported by the DIM ACAV and the French National Agency for Research (ANR-13-JS05-0005-01). SL acknowledges fruitfull discussions on the subject with Dr. Micheal Ireland and Dr. Herv\'e Le Coroller. The authors would also like to thank D. Buscher, the referee, for his insightful comments and contribution to Section~\ref{sc:basic}.

\bibliographystyle{aa}   %>>>> makes bibtex use aa.bst
\bibliography{Nulbib}

\onecolumn 

\appendix

\section[]{Numerical version of the V2PM of 2 IOs 4 beams combiner}
\label{appen}
$V2PM_a$ and $V2PM_b$ can be calculated from theory according to the mapping
presented in Figure~\ref{fig:IO}. $V2PM_b$ is a linear combination of $V2PM_a$. Assuming 33/33/33 tri-couplers, and
perfect ABCD ($0$,$\pi/2$,$\pi$,$3\pi/2$), they wrote:
\begin{equation}
V2PM_a =
\begin{pmatrix}
\oh & 0 & 0 & 0  & 0 & 0 & 0 & 0 & 0 & 0 \\
\os &\os& 0 & 0  &\op& 0 & 0 & 0 & 0 & 0 \\
\os &\os& 0 & 0  &\om& 0 & 0 & 0 & 0 & 0 \\
\ov &\oz&\ov& 0  &\oa+\oa&\oa& 0 &\oa+\oa& 0 & 0 \\
\ov & 0 &\ov& 0  &\oa+\ob&\ob& 0 &\oa+\ob& 0 & 0  \\
\ov &\ow&\ov& 0  &\oa+\oc&\oc& 0 &\oa+\oc& 0 & 0 \\
\ov &\ow&\ov& 0  &\oa+\od&\od& 0 &\oa+\od& 0 & 0 \\
0   &\os&\os& 0  & 0 & 0 & 0 &\op& 0 & 0 \\
\ov &\ov&\ov&\ov &\oa&\oa&\oa&\oa&\oa&\oa\\
\ov &\ov&\ov&\ov &\oa&\ob&\ob&\ob&\ob&\oa\\
\ov &\ov&\ov&\ov &\oa&\oc&\oc&\oc&\oc&\oa\\
\ov &\ov&\ov&\ov &\oa&\od&\od&\od&\od&\oa\\
0   &\os&\os& 0  & 0 & 0 & 0 &\om& 0 & 0 \\
0   &\ov&\oz&\ov & 0 & 0 & 0 &\oa+\oa&\oa&\oa+\oa\\
0   &\ov& 0 &\ov & 0 & 0 & 0 &\oa+\ob&\ob&\oa+\ob\\
0   &\ov&\ow&\ov & 0 & 0 & 0 &\oa+\oc&\oc&\oa+\oc\\
0   &\ov&\ow&\ov & 0 & 0 & 0 &\oa+\od&\od&\oa+\od\\
0   & 0 &\os&\os & 0 & 0 & 0 & 0 & 0 &\op\\
0   & 0 &\os&\os & 0 & 0 & 0 & 0 & 0 &\om\\
0   & 0 & 0 &\oh & 0 & 0 & 0 & 0 & 0 & 0 \\
\end{pmatrix}\label{eq:V2PMa}
\end{equation}
and
\begin{equation}
V2PM_b =
\begin {pmatrix}
\oh & 0 & 0 & 0 & 0 & 0 & 0 & 0 & 0 & 0 \\
\os &\os& 0 & 0 &\op& 0 & 0 & 0 & 0 & 0 \\
\os &\os& 0 & 0 &\om& 0 & 0 & 0 & 0 & 0 \\
\ov &\ow&\ov& 0 &\oa&\oa& 0 &\oa& 0 & 0 \\
\ov &\ow&\ov& 0 &\oa&\oa& 0 &\ob& 0 & 0 \\
\ov &\ow&\ov& 0 &\oa&\oa& 0 &\oc& 0 & 0 \\
\ov &\ow&\ov& 0 &\oa&\oa& 0 &\od& 0 & 0 \\
0   &\os&\os& 0 & 0 &\op& 0 & 0 & 0 & 0 \\
\ov &\ov&\ov&\ov&\oa& 0 &\oa& 0 &\oa& 0 \\
\ov &\ov&\ov&\ov&\oa& 0 &\oa& 0 &\ob& 0 \\
\ov &\ov&\ov&\ov&\oa& 0 &\oa& 0 &\oc& 0 \\
\ov &\ov&\ov&\ov&\oa& 0 &\oa& 0 &\od& 0 \\
0   &\os&\os& 0 & 0 &\om& 0 & 0 & 0 & 0 \\
0   &\ov&\ow&\ov& 0 &\oa&\oa& 0 & 0 &\oa\\
0   &\ov&\ow&\ov& 0 &\oa&\oa& 0 & 0 &\ob\\
0   &\ov&\ow&\ov& 0 &\oa&\oa& 0 & 0 &\oc\\
0   &\ov&\ow&\ov& 0 &\oa&\oa& 0 & 0 &\od\\
0   & 0 &\os&\os& 0 & 0 &\op& 0 & 0 & 0 \\
0   & 0 &\os&\os& 0 & 0 &\om& 0 & 0 & 0 \\
0   & 0 & 0 &\oh& 0 & 0 & 0 & 0 & 0 & 0 \\
\end{pmatrix}\label{eq:V2PMb}
\end{equation}

On the other hand, the theoretical $V2PM$ of the GRAVITY instrument will be
\citep[from][]{2008SPIE.7013E..31L}:
\begin{equation}
V2PM_{\rm GRAVITY}
=    
\left(
\begin {array}{cccccccccc}
\gop & \gop & 0 & 0 & \ggoa & 0 & 0 & 0 & 0 & 0  \\
\gop & \gop & 0 & 0 & \gob & 0 & 0 & 0 & 0 & 0 \\
\gop & \gop & 0 & 0 & \goc & 0 & 0 & 0 & 0 & 0 \\
\gop & \gop & 0 & 0 & \god & 0 & 0 & 0 & 0 & 0 \\
 0 &\gop & \gop & 0 & 0 & \ggoa & 0 & 0 & 0 & 0  \\
 0 &\gop & \gop & 0 & 0 & \gob & 0 & 0 & 0 & 0  \\
\gop &  0 &\gop & 0 & 0 & 0 & 0 & 0 & 0 & \ggoa  \\
\gop &  0 &\gop & 0 & 0 & 0 & 0 & 0 & 0 & \gob  \\
\gop & 0 & \gop & 0 & 0 & 0 & 0 & 0 & 0 & \goc  \\
\gop & 0 & \gop & 0 & 0 & 0 & 0 & 0 & 0 & \god  \\
\gop & 0 & 0 & \gop & 0 & 0 & 0 & \ggoa & 0 & 0 \\
\gop & 0 & 0 & \gop & 0 & 0 & 0 &  \gob & 0 & 0  \\
\gop & 0 & 0 & \gop & 0 & 0 & 0 &  \goc & 0 & 0  \\
\gop & 0 & 0 & \gop & 0 & 0 & 0 &  \god & 0 & 0  \\
 0 &\gop & 0 & \gop & 0 & 0 & \ggoa & 0 & 0 & 0  \\
 0 &\gop & 0 & \gop & 0 & 0 & \gob & 0 & 0 & 0  \\
 0 &\gop & 0 & \gop & 0 & 0 & \goc & 0 & 0 & 0   \\
 0 &\gop & 0 & \gop & 0 & 0 & \god & 0 & 0 & 0   \\
 0 &\gop & \gop & 0 & 0 & \goc & 0 & 0 & 0 & 0  \\
 0 &\gop & \gop & 0 & 0 & \god & 0 & 0 & 0 & 0  \\
 0 & 0 &\gop & \gop & 0 & 0 & 0 & 0 & \ggoa & 0  \\
 0 & 0 &\gop & \gop & 0 & 0 & 0 & 0 & \gob & 0  \\
 0 & 0 &\gop & \gop & 0 & 0 & 0 & 0 & \goc & 0  \\
 0 & 0 &\gop & \gop & 0 & 0 & 0 & 0 & \god & 0  
\end {array}
\right)
\label{GRAVITY}
\end{equation}

\label{lastpage}
\end{document}